\documentclass[dvips]{arxstspdf}
\usepackage{graphicx}
\usepackage{flushend,stfloats}

\volume{25}
\issue{3}
\pubyear{2010}
\firstpage{348}
\lastpage{367}
\doi{10.1214/10-STS329}

\makeatletter
\setattribute{abstract}    {skip} {20\p@}
\setattribute{keyword}     {skip} {8\p@}
\setattribute{frontmatter} {skip} {0\p@ plus 3\p@ minus 3\p@}
\setattribute{frontmatter} {cmd}  {}
\setattribute{abstract}   {width}  {36.5pc}
\setattribute{keyword}    {width}  {36.5pc}

\newcommand{\dsum}{\sum}
\newcommand{\dint}{\int}
\newcommand{\m}{\mathbf{m}}
\newcommand{\z}{\mathbf{z}}
\newcommand{\Y}{\mathbf{Y}}
\newcommand{\W}{\mathbf{W}}
\newcommand{\Q}{\mathbf{Q}}
\newcommand{\bet}{\bolds{\beta}}
\newcommand{\Thet}{\bolds{\Theta}}
\newcommand{\eps}{\bolds{\epsilon}}

\newtheorem{Theorem}{Theorem}
\newtheorem{Proposition}{Proposition}
\newproclaim{Example}{Example}
\newtheorem{Lemma}{Lemma}

\makeatother

\begin{document}
\begin{frontmatter}

\title{On the Sample Information About Parameter and Prediction}
\runtitle{Information About Parameter and Prediction}

\begin{aug}
\author[a]{\fnms{Nader} \snm{Ebrahimi}\ead[label=e1]{nader@math.niu.edu}},
\author[b]{\fnms{Ehsan S.} \snm{Soofi}\corref{}\thanksref{t1}\ead[label=e2]{esoofi@uwm.edu}}
\and
\author[c]{\fnms{Refik} \snm{Soyer}\ead[label=e3]{soyer@gwu.edu}}
\runauthor{N. Ebrahimi, E. S. Soofi and R. Soyer}
\thankstext{t1}{Corresponding author.}

\address[a]{Nader Ebrahimi is Professor,
Division of Statistics,
Northern Illinois University,
DeKalb, Illinois 60155, USA (\printead{e1}).}
\address[b]{Ehsan S. Soofi is Professor of Management Science and Statistics,
Sheldon B. Lubar School of Business, University of Wisconsin-Milwaukee,
PO Box 742, Milwaukee, Wisconsin 53201, USA (\printead{e2}).}
\address[c]{Refik Soyer is Professor of Decision Sciences and Statistics,
Department  of Decision Sciences and Department  of Statistics,
George Washington University, Washington, DC 20052, USA (\printead{e3}).}

\end{aug}

%
\begin{abstract}
The Bayesian measure of sample information about the parameter, known
as Lindley's measure,
is widely used in various problems
such as developing prior distributions, models for the likelihood
functions and optimal designs.
The predictive information is defined similarly and used for model
selection and optimal
designs, though to a lesser extent.
The parameter and predictive information measures are proper utility
functions and
have been also used in combination.
Yet the relationship between the two measures and the effects of conditional
dependence between the observable quantities on the Bayesian
information measures remain unexplored.
We address both issues. The relationship between the two information
measures is explored
through the information provided by the sample about the parameter and
prediction jointly.
The role of dependence is explored along with the interplay between the
information measures, prior and sampling design.
For the conditionally independent sequence of observable quantities,
decompositions of the joint
information characterize Lindley's measure as the sample information
about the parameter and prediction
jointly and the predictive information as part of it.
For the conditionally dependent case, the joint information about
parameter and prediction
exceeds Lindley's measure by an amount due to the dependence.
More specific results are shown for the normal linear models and a
broad subfamily of the
exponential family. Conditionally independent samples provide
relatively little information
for prediction, and the gap between the parameter and predictive
information measures
grows rapidly with the sample size.
Three dependence structures are studied: the intraclass (IC) and
serially correlated (SC) normal
models, and order statistics. For IC and SC models, the information
about the mean parameter
decreases and the predictive information increases with the
correlation, but the joint information
is not monotone and has a unique minimum. Compensation of
the loss of parameter information due to dependence requires larger samples.
For the order statistics, the joint information exceeds
Lindley's measure by an amount which does not depend on the prior or
the model for the data,
but it is not monotone in the sample size and has a unique maximum.
\end{abstract}

\dedicated{\textit{The authors dedicate this article to Professor Dennis V.
Lindley in appreciation of his insightful pioneering work on the
relationships between Shannon's information theory and Bayesian inference.}}

%
\begin{keyword}
\kwd{Bayesian predictive distribution}
\kwd{entropy}
\kwd{mutual information}
\kwd{optimal design}
\kwd{reference prior}
\kwd{intraclass correlation}
\kwd{serial correlation}
\kwd{order statistics}.
\end{keyword}

\end{frontmatter}

\section{Introduction}

The elements of Bayesian information analysis are a set of $n$
observations, denoted as an $n \times1$
vector $\mathbf y$ generated from a sequence of random variables $Y_1, Y_2,
\ldots$ with a joint probability
model $f(\mathbf y|\theta)$ where the parameter $\theta$ has a prior
probability distribution
$f(\theta), \theta\in\Theta$ and a new outcome $Y_\nu$.
We follow the convention of using uppercase letters for unknown
quantities, which may be scalar or vector.
Whereas the concept of prediction is usually an afterthought in
classical statistics,
unless one deals with regression or forecasting type models,
predictive inference naturally arises as a consequence of calculus of
probability
and is a standard output of Bayesian analysis.
Bayesians are interested in prediction of future outcomes, because
eventually they
will be observed and allow to settle bets in the sense of de Finetti.
The predictive inference is considered as a distinguishing feature of
the Bayesian approach.
But one cannot develop predictive inference without estimation,
that is, without obtaining the posterior distribution of the parameter.
The parameter
plays the pivotal role in prediction, and a clear perspective of the
information provided by the sample
about the parameter and prediction can be obtained only through viewing
$(\Theta, Y_\nu)$ jointly.

Information provided by the data refers to a measure that quantifies
changes from a prior to a posterior
distribution of an unknown quantity.
Lindley (\citeyear{1956Lindley}) framed the problem of measuring sample information about
the parameter
in terms of Shannon's (\citeyear{1948Shannon}) notion of information in the noisy channel
(sample) about the
signal transmitted from a source (parameter). The notion is
operationalized in terms
of entropy and mutual information measures.
Bernardo (\citeyear{1979aBernardo}) showed that Lindley's measure of information about the
parameter
is the expected value of a logarithmic utility function
for the decision problem of reporting a probability distribution
from the space of all distributions.
The information utility function belongs to a large class of utility functions
discussed by Good (\citeyear{1971Good}) and others which lead to the posterior
distribution given by the Bayes rule
as the optimal distribution.
The predictive version of Lindley's measure, referred to as predictive
information,
quantifies the expected amount of information provided by the sample
about prediction of
a new outcome.

A list of articles on Lindley's measure and its methodological
applications is tabulated in the\break \hyperref[app]{Appendix}.
The major areas of applications are classified in terms of sampling
design and developing
models for the likelihood function, and developing prior and posterior
distributions.
Stone (\citeyear{1959Stone}) was first to apply Lindley's measure to the normal
regression experiments and
El-Sayyed (\citeyear{1969El}) was first to apply Lindley's measure to the
exponential model.
Following Bernardo (\citeyear{1979aBernardo}, \citeyear{1979bBernardo}), several authors have presented evaluation
and selection of the likelihood
function in terms of Lindley's measure as a Bayesian decision problem.
Chaloner and Verdinelli (\citeyear{1995Chaloner}) provided an extensive review and
additional references
for the experimental design; see also the works of Barlow and Hsiung
(\citeyear{1983Barlow}) and Polson (\citeyear{1993Polson}).
Soofi (\citeyear{1988Soofi}, \citeyear{1990Soofi}) and Ebrahimi and Soofi (\citeyear{1990Ebrahimi}) examined the trade-offs
between the prior and design parameters for the information about the
model parameter.
Carota, {Parmigiani} and {Polson} (\citeyear{1996Carota}) developed an approximation for application to
model elaboration.
Yuan and Clarke (\citeyear{1999Yuan}) proposed developing the model for the
likelihood function that maximizes Lindley's measure subject to
a constraint in terms of the Bayes risk of the model.
San Martini and Spezzaferri (\citeyear{1984San}) used a version of the predictive
information for model selection.
Amaral and Dunsmore (\citeyear{1985Amaral}) studied the predictive measure and applied
it to the exponential
parameter. Verdinelli, {Polson} and {Singpurwalla} (\citeyear{1993Verdinelli}) used the predictive information and
Verdinelli (\citeyear{1992Verdinelli}) considered a linear combination of the parameter and
predictive information measures
as design criteria.

This article is another testimony of the depth and breadth of Lindley's
pioneering work on the relationships
between Shannon's information theory and\break Bayesian inference.
We explore the relationship between the parameter and predictive
information measures and examine
the roles of prior, design and the dependence in the sequence $Y_i |
\theta,  i=1,2, \ldots,$
on the information measures and their interrelationship.
This expedition integrates and expands the existing literature in three
directions.

\begin{figure}
\begin{tabular}{c}

\includegraphics{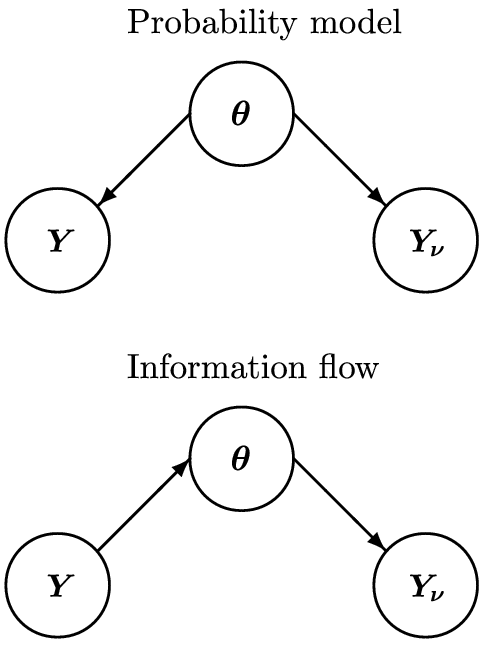}
\\
(a) Conditional independent\\[6pt]

\includegraphics{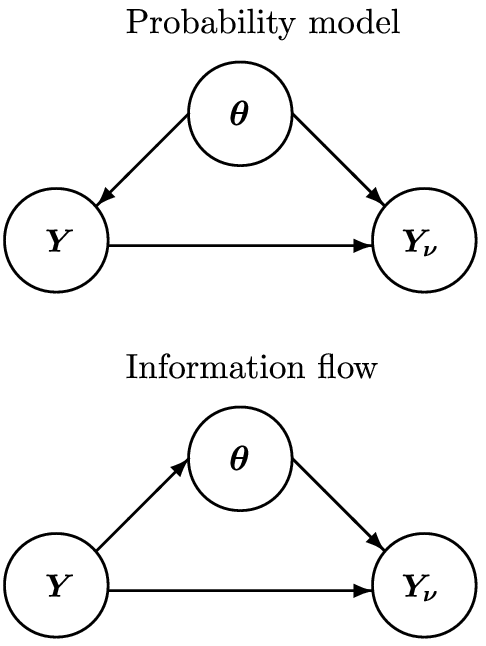}
\\
(b) Conditional dependent
\end{tabular}
\caption{Graphics of conditional independent and
 dependent models. \textup{(a)} Conditional independent. \textup{(b)} Conditional dependent.}
\label{fig1}
\end{figure}

First, to this date, the relationship between the sample information
about the parameter (Lindley's measure)
and predictive information remains unexplored.
Lindley's measure focuses on the information flow between the pair $(\Y
, \Theta)$.
The predictive information measure is based on the information flow
between the pair $(\Y, Y_\nu)$.
The key to exploring the relationship between the information provided
by the sample
about the parameter and for the prediction is through viewing $(\Theta
, Y_\nu)$ jointly as
an interrelated pair. In this perspective, $\Theta$ plays an
intermediary role in the information flow from
the data $\mathbf y$ to the prediction quantity $Y_\nu$. The information
flow from $\Y$ to
the pair $(\Y, Y_\nu)$ is different
when $Y_i | \theta$,  $i=1,2, \ldots,$ are conditionally independent
and conditionally dependent.
Panel (a) of Figure~\ref{fig1} depicts the conditionally independent model and
its information flow diagram.
In this case, the parameter $\theta$ is the only link between $\Y$
and $Y_\nu$, thus
the information flows from the data to the predictive distribution
solely through the parameter.
This information flow from $\Y$ to $\Theta$ to $Y_\nu$ is analogous
to the data processing
of the information theory
(Cover and Thomas, \citeyear{1991Cover}) where $(\Y, \Theta, Y_\nu)$ is a Markovian triplet.
We will show that in this case the sample information about the
parameter is in fact the
entire information provided by $\Y$
about $(\Theta, Y_\nu)$ jointly, and that the predictive information
is only a part of it.
We will further show that for some important classes of models, such as
the normal linear model
and a large family of lifetime models,
the predictive information provided by the conditionally independent
sample is only
a small fraction of the parameter (joint) information.

Second, thus far, the effects of dependence in the sequence
$Y_i | \theta, \ i=1,2, \ldots,$ on the Bayesian information measures
remain unexplored.
Panel (b) of Figure~\ref{fig1} shows the graphical representations of the
conditionally dependent model
and its information flow diagram. In this case,
the information flows from the data to predictive
distribution directly due to the conditional dependence, as well as
indirectly via the parameter.
Consequently, the relationship between the parameter and predictive
information measures is
quite different than that for the conditionally independent case.
We will show that for the conditionally dependent case, the sample
information for the pair $(\Theta, Y_\nu)$
decomposes into the information about the parameter (Lindley's measure)
and an information measure
mapping the conditional dependence.
We study the role of dependence for three important cases: the
intraclass (IC) and serial correlation (SC)
dependence structures for the normal sample, and order statistics where
no particular distribution
is specified for the likelihood and prior.
Estimation of the normal mean and prediction under the IC and SC models
are commonplace.
We examine the effects of dependence on the parameter and predictive
information measures drawing
from Pourahmadi and Soofi's (\citeyear{2000Pourahmadi}) study of information\vadjust{\goodbreak} measures for
prediction of future outcomes
in time series.
We will show that the sample can provide a substantial amount of
information for prediction
and the dominance of parameter information that was noted for the
conditionally independent case no longer holds.
Order statistics, which conditional on the parameter form a Markovian
sequence (Arnold, {Balakrishnan} and {Nagaraja}, \citeyear{1992Arnold}),
also provide a useful context for studying the effects of dependence on
information measures.
For example, in life testing, the information that the first $r$
failure times provide
about the model parameter as well as about the time
to next failure $Y_{r+1}$ are of interest. Here, $n$ items are under
the test,
failures are observed one at a time, and it is desirable to determine
at an early
stage how costly the testing is
going to be and whether an action such as a redesign is warranted.
Such joint parameter--predictive inferences were considered by Lawless
(\citeyear{1971Lawless}), Kaminsky
and Rhodin (\citeyear{1985Kaminsky}) and Ebrahimi (\citeyear{1992Ebrahimi}) under various sampling plans.

Third, the Bayesian information research has focused either on the
design or on the prior.
The past research has mainly used two types of models encompassing two
different parameters:
the linear model for the normal mean parameter, and the lifetime model
where the scale parameter
of an exponential family distribution is of interest.
We consider the normal linear model with normal prior distribution for
the mean and
a subfamily of the exponential family under the gamma prior
distribution for the scale parameter.
This subfamily includes the exponential
distribution and many parametric families such as Weibull, Pareto and Gumbel
extreme value. For each class of models, we examine the relationships
between the parameter
and predictive information measures. Furthermore, we explore the
effects of sampling plan
and prior distribution on the parameter and predictive information measures.
We will show that under the optimal design for the parameter estimation,
the loss of information for prediction
is not nearly as severe as the loss of information about the parameter
under the optimal design for prediction.

This article is organized as follows.
Section \ref{IM} pre\-sents the measures of information provided by the
sample about the
parameter and prediction, including results on the relationship between
them for the conditionally
independent model.
Section \ref{LM} explores the measures of information provided by the
sample about the
parameter and prediction in terms of the prior and design matrix for
linear models.
Section \ref{EXPO} explores the measures of information provided by
the sample about the
parameter and prediction for a subfamily of the exponential family and
explores the
interplay between parameter and predictive information for a broad
family of
distributions generated by transformations of the  exponential model.
Section \ref{DEPEND} examines information measures for conditionally
dependent samples.
Section~\ref{CONCL} gives the concluding remarks. The \hyperref[app]{Appendix} provides
a classification
of the literature on Bayesian applications of the mutual information
and some
technical details.

\section{Information Measures} \label{IM}

Let $Q$ represent the unknown quantity of interest: $\Theta, \ Y_\nu
$, individually or as a pair, or a function of them.
For notational convenience we represent probability distribution with
its density function $f(\cdot)$ and
use subscript $i$ for the elements of data vector $\mathbf y$ and $Y_\nu,
\ \nu\neq i$, for prediction.
Information provided by the data $\mathbf y$ about $Q$ is measured by a
function that maps changes between a prior
distribution $f(q)$ and the posterior distribution $f(q|\mathbf y)$
obtained via the Bayes rule.
Two measures of changes of the prior and posterior distributions are as follows.
The uncertainty about $Q$ is measured by Shannon entropy
\[
H(Q)=H(f)= - \dint f(q) \log f(q)\, dq,
\]
and the observed sample information about $Q$ is measured by the
entropy difference
\begin{equation}
\label{DHQ}
\Delta H(\mathbf y; Q) = H(Q) - H(Q| \mathbf y).
\end{equation}
The information discrepancy between the prior and posterior
distributions is measured by the Kullback--Leibler divergence
\begin{equation}
\label{K}
\quad K[f(q|\mathbf y)\dvtx f(q)]= \dint f(q|\mathbf y) \log\frac{f(q|\mathbf y)}{f(q)} \,dq
\geq0,\hspace*{-12pt}
\end{equation}
where the equality in (\ref{K}) holds if and only if $f(q|\mathbf y)=f(q)$
almost everywhere.
The observed sample information measure (\ref{DHQ}) can be positive or
negative depending on which of the two distributions is more
concentrated (less uniform).
For a $k$-dimensional random \mbox{vector} $\Q$, an orthonormal $k \times k$
matrix $A$
and a $k \times1$ vector $\mathbf c$, $H(A \Q+\mathbf c)=H(\Q)$, but (\ref{DHQ})
is invariant under all linear transformations of $\Q$.
The information discrepancy (\ref{K}) is a relative entropy which only
detects changes between
the prior and the posterior, without indicating which of the two
distributions is more informative.
It is invariant under all one-to-one transformations of $Q$.

The expected sample information measures are obtained by viewing the
observed information measures
(\ref{DHQ}) and (\ref{K}) as functions of the data and averaging them
with respect
to the marginal distribution of $\Y$.
The expected entropy difference and expected Kullback--Leibler
divergence provide the
same measure, known as the \textit{mutual information}
\begin{eqnarray}
\label{M}
M(\Y; Q) &=& E_{\mathbf y} \{\Delta H(\mathbf y; Q) \}\nonumber
\\[-8pt]\\[-8pt]
&=& E_{\mathbf y} \{K[f(q|\mathbf
y)\dvtx f(q)] \},\nonumber
\end{eqnarray}
where $E_{\mathbf y}$ denotes averaging with respect to
\[
f(\mathbf y) = \dint f(\theta) f(\mathbf y|\theta)\,d \theta.
\]
Other representations of $M(\Y; Q)$ are
\begin{eqnarray}
\label{M2}
M(\Y; Q) & = & H(Q)- \mathcal{H}(Q|\Y)\nonumber
\\[-8pt]\\[-8pt]
&= &K[f(q,\mathbf y)\dvtx f(q) f(\mathbf y)] ,\nonumber
\end{eqnarray}
where
\[
\mathcal{H}(Q|\Y)= E_{\mathbf y}\{H(Q| \mathbf y) \} = \dint H(Q| \mathbf y) f(\mathbf
y)\,d\mathbf y
\]
is referred to as the \textit{conditional entropy} in the information
theory literature.
The first representations in (\ref{M}) and (\ref{M2}) are in terms of
the expected
uncertainty reduction, and the second representation in (\ref{M2})
shows that the
mutual information is symmetric in $Q$ and $\Y$.
It is noteworthy to mention that the equalities in (\ref{M}) and (\ref
{M2}) do not hold,
in general, for generalizations of Shannon entropy and Kullback--Leibler
information divergence, such as R\' enyi measures; see the article by
Ebrahimi, {Soofi} and {Soyer} (\citeyear{2010Ebrahimi}).

Some useful properties of the mutual information are as follows:
\begin{enumerate}[1.]
\item
$M(\Y; Q)\geq0$, where the equality holds if and only if $Q$ and $\Y
$ are independent.
\item
The conditional mutual information is defined by $M(\Y; Q |S) =
E_s[M(\Y; Q |s)] \geq0$,
where the\break equality holds if and only if $Q$ and $\Y$ are conditionally
independent.
\item
Given $f(q)$, $M(\Y; Q)$ is convex in $f(q|\mathbf y)$ and given $f(q|\mathbf
y)$, $M(\Y; Q)$ is concave in $f(q)$.
\item
Let $\mathbf{Y}_n$ denote a vector of dimension $n, \ Y_j \in\mathbf{Y}_n$
and $Y_j \notin\mathbf{Y}_{n-1}$. Then
\begin{eqnarray}
\label{MMMM}
\quad M(\mathbf{Y}_n; Q) &=& M(\mathbf{Y}_{n-1}; Q)+M(Q; Y_j |\mathbf{Y}_{n-1})\nonumber
\\[-8pt]\\[-8pt]
&\geq&
M(\mathbf{Y}_{n-1}; Q),\nonumber
\end{eqnarray}
thus $M(\mathbf{Y}_n; Q)$ is increasing in $n$.
\item
$M(\Y;Q)$ is invariant under one-to-one transformations of $Q$ and $\Y$.
\end{enumerate}

\subsection{Marginal Information}

For $Q=\Theta$, the observation $\mathbf y$ provides the likelihood function,
$\mathcal{L} (\theta) \propto f(\mathbf y|\theta)$ and updates the prior to
the posterior distribution
\begin{equation}
\label{POSTF}
f(\theta|\mathbf y) \propto f(\theta) f(\mathbf y|\theta).
\end{equation}
The expected sample information about the parameter,
$M(\Y; \Theta)$, is known as Lindley's measure (Lindley, \citeyear{1956Lindley})
and is referred to as the parameter information.

The following properties are also well known:
\begin{enumerate}[1.]
\item
Let $S_n=S(\Y)$ be a general transformation. Then $M(\Y; \Theta)
\geq M(S_n; \Theta)$, where the equality holds if and
only if $S_n$ is a sufficient statistic for $\theta$.
\item
$M(\mathbf{Y}_n; \Theta)$ is concave in $n$, which implies that $M(Y_j;
\Theta|\mathbf{Y}_{n-1}) \leq M(Y_j; \Theta)$.
\item
Ignorance between two neighboring values in the parameter space,
$P(\theta)= P(\theta+ \delta(\theta))=0.5$, implies that
$M(\Y; \Theta) \approx2 \delta^2 (\theta) \mathcal{I}_F(\theta)$ as
$\delta\theta\to0$, where $\mathcal{I}_F(\theta)$
is Fisher information (Lindley, \citeyear{1961Lindley}, page 467).
Similar approximation holds more generally for $M(\Y; Q)$; see the
classic book of Kullback (\citeyear{1959Kullback}).

\end{enumerate}

For $Q=Y_\nu$, the prior and posterior predictive distributions,
respectively, are given by
\[
f(y_\nu) = \dint f(y_\nu|\theta)f(\theta)\, d \theta
\]
and
\begin{equation}
\label{PREDF}
f(y_\nu|\mathbf y) = \dint f(y_\nu|\theta) f(\theta|\mathbf y)\, d \theta.
\end{equation}
The expected information $M(\Y; Y_\nu)$ is referred to as the
predictive information
(San Martini and Spezzaferri, \citeyear{1984San}; Amaral and Dunsmore, \citeyear{1985Amaral}).

In some problems, both the parameter and the prediction are of interest
(Chaloner and Verdinelli, \citeyear{1995Chaloner}).
Verdinelli (\citeyear{1992Verdinelli}) proposed the linear combination of marginal utilities
\begin{equation}
\label{UMM}
U(\Y; \Thet,Y_\nu) =w_1 M(\Y; \Thet)+w_2M(\Y;Y_\nu),
\end{equation}
where $w_k \geq0,  k=1,2,$
are weights that reflect the relative importance of the parameter and
prediction for the experimenter.
Since $\Thet$ and $Y_\nu$ are not independent quantities, $M(\Y;
\Thet)$ and $M(\Y;Y_\nu)$ are
not additively separable. The weights in (\ref{UMM}) do not take into
account the dependence between the
prediction and the parameter.

\subsection{Joint Information}

Taking the dependence between the parameter and prediction into account
requires considering the joint information
for the vector of parameter and prediction.
The observed and expected information measures are defined by (\ref
{DHQ}) and (\ref{M}) where
$Q=(\Theta, Y_\nu)$, and will be denoted as $\Delta H[\mathbf y; (\Theta
, Y_\nu)]$ and $M[\Y;\break (\Theta, Y_\nu)]$.
The next theorem encapsulates the relationships between the joint,
parameter and predictive information measures
for the conditionally independent samples.

\begin{Theorem}  \label{BDPI}
If $Y_1|\theta, Y_2|\theta, \ldots$ are conditionally independent, then:
\begin{longlist}[(a)]
\item[(a)]
$\Delta H(\mathbf y; \Theta)=\Delta H[\mathbf y; (\Theta, Y_\nu)]$;
\item[(b)]
$M(\Y; \Theta) = M[\Y; (\Theta, Y_\nu)]$;
\item[(c)]
$M(\Y; Y_\nu) \leq M(\Y; \Theta)$.
\end{longlist}
\end{Theorem}

\begin{pf}
The proof of (a) is as follows. The joint entropy decomposes additively as
\[
H(\Theta, Y_\nu) = H(\Theta) +\mathcal{H}(Y_\nu|\Theta),
\]
where $\mathcal{H}(Y_\nu|\Theta)=E_\theta\{ H(Y_\nu|\theta)\}$ is the
conditional
entropy. Letting $Q=(\Theta, Y_\nu)$ in (\ref{DHQ})
and applying the entropy decomposition to each entropy, we have
\begin{eqnarray*}
\Delta H[\mathbf y; (\Theta, Y_\nu)] &=& H(\Theta) +\mathcal{H}(Y_\nu|\Theta
)
\\
&&{}-\{H(\Theta|\mathbf y) + \mathcal{H}(Y_\nu|\Theta, \mathbf y)\},
\end{eqnarray*}
where $\mathcal{H}(Y_\nu|\Theta, \mathbf y)=E_\theta\{ H(Y_\nu|\theta,
\mathbf y)\}$.
The first and third terms give $\Delta H(\mathbf y; \Theta)$. The
conditional independence implies for each
$\theta,  H[f(y_\nu|\theta, \mathbf y)]=H[f(y_\nu|\break \theta)]$, thus
$E_\theta\{ H(Y_\nu|\theta, \mathbf y)\}=E_\theta\{ H(Y_\nu|\theta)\}
$, and the second and fourth
terms cancel out, which gives (a). Since $\Y\to\Theta\to Y_\nu$ is
a Markovian triplet,
parts (b) and (c) are implied by properties of the mutual information
functions of Markovian sequences (see, e.g., Cover and Thomas, \citeyear{1991Cover},
pages 27, 32--33).
\end{pf}

By part (a) of Theorem \ref{BDPI}, under the conditionally independent model,
the information provided by each and every sample about the parameter
is the same as
the joint information for the parameter and prediction.

Part (b) of Theorem \ref{BDPI} provides a broader interpretation of
Lindley's information, namely expected information
provided by the data about the parameter and for the prediction. An
immediate implication is
that the prior distribution (Bernardo, \citeyear{1979aBernardo}, \citeyear{1979bBernardo}), the design (Chaloner
and Verdinelli, \citeyear{1995Chaloner}; Polson, \citeyear{1993Polson}) and the likelihood model (Yuan and
Clarke, \citeyear{1999Yuan}) that maximize $M(\Y; \Theta)$
also maximize sample information about the parameter and prediction jointly.
However, by part (c) of Theorem \ref{BDPI}, such optimal prior,
design, and model may not be
optimal according to $M(\Y; Y_\nu)$. Similarly, the optimal design of
Verdinelli, {Polson} and {Singpurwalla} (\citeyear{1993Verdinelli})
and the optimal model of San Martini and Spezzaferri (\citeyear{1984San}) which
maximize $M(\Y; Y_\nu)$
may not be optimal according to $M(\Y; \Theta)$.

The inequality in (c)
is the Bayesian version of the information processing inequality of
information theory,
and can be referred to as the \textit{Bayesian data processing inequality} mapping
the information flow $\Y\to\Theta\to Y_\nu$ through (\ref{POSTF})
and (\ref{PREDF}),
as shown in Figure~\ref{fig1}(a).

By part (b) of Theorem \ref{BDPI} and decomposition of $M[\mathbf{Y};(\Theta,Y_\nu)]$ we have
\begin{equation}
\label{MMM1}
M(\Y;\Theta) = M (\Y; Y_\nu) + M(\Y; \Theta|Y_\nu) ,
\end{equation}
where $M[(\Y;\Theta) |Y_\nu]= E_{y_\nu} \{K[f(\mathbf y, \theta)|y_\nu
)\dvtx f(\theta|\break y_\nu)f(\mathbf y|y_\nu)] \}$
is the conditional mutual information \mbox{between} $\Theta$ and $\Y$,
given $Y_\nu$.
This measure is the link between the parameter and predictive
information measures and
is key for studying their relationship.
Applying (\ref{MMM1}) to the utility function (\ref{UMM}) gives the
weights for the additive information
measures in (\ref{MMM1}) as
\begin{eqnarray}
\label{UMM2}
&&U(\Y; \Thet,Y_\nu)\nonumber
\\[-8pt]\\[-8pt]
&&\quad= w_1 M(\Y; \Thet|Y_\nu)+ (w_1+w_2) M(\Y
;Y_\nu).\nonumber
\end{eqnarray}

\section{Linear Models} \label{LM}

Consider the normal linear model
\[
\mathbf y= X \bet+ \eps,
\]
where $\mathbf y$ is an $n\times1$ vector of observations,
$X$ is an $n\times p$ design matrix, $\bet$ is the $p \times1$
parameter vector,
$\eps$~is the error vector. Under the conditionally independent model
$f(\eps|\bet)=N(\mathbf0, \sigma^2_1 I_n)$,
$\sigma^2_1>0$ is known and $I_n$ is identity matrix of dimension $n$.

It will be more insightful to use the orthonormal rotation $Z=XG$
and $\bolds{\theta}=G^\prime\bet$, where $G$ is the matrix
of eigenvectors of $X^\prime X$, and $\Lambda= Z^\prime Z =
\operatorname{diag} [ \lambda_1, \ldots, \lambda_p]$
where $\lambda_j >0,  j=1, \ldots, p$, are the eigenvalues of
$X^\prime X$.
By the invariance of entropy\break under orthonormal transformations,
$\Delta H(\mathbf y; \Thet)=\break \Delta H(\mathbf y; \bet)$ and
by invariance of mutual information under all one-to-one
transformations, $M(\Y; \Thet)=M(\Y; \bet)$.

We use the normal conjugate prior $f(\bolds{\theta})=N(\m_0,\break \sigma^2_0 V_0)$,
where $V_0 = \operatorname{diag} [ v_{01}, \ldots, v_{0p}]$.
The posterior distribution is $f(\bolds{\theta}|\mathbf y)=N(\m_1, \sigma
^2_1 V_1)$ where
$\m_1= V_1^{-1} ( \eta V_0^{-1} \m_0 + Z^\prime\mathbf y ),
V_1= (\eta V_0^{-1}+Z^\prime Z )^{-1}$ and $\eta= \frac{\sigma
^2_1}{\sigma^2_0}$.
All distributions and informationc\vspace*{-2pt} measures\break are~conditional on $Z$ and
$\sigma^2_1$ which are assumed
to be given. The prior and posterior entropies are
$H(\Thet|\sigma^2_k V_k)=\frac{p}{2}\log(2 \pi e) + \frac{1}{2}
\log|\sigma^2_k V_k|,  k=0,1$,
where $|\cdot|$ denotes the determinant.
Since entropy is location invariant, $\m_k$ does not matter. Also
since $V_1$ does not depend on data $\mathbf y$, the conditional entropy
and posterior entropies are equal,
$\mathcal{H}(\Thet|\Y, Z, \eta, V_0)=H(\Thet|\mathbf y, Z, \eta, V_0)$.
Thus, the observed and expected
sample information measures are the same, given by
\begin{eqnarray}
\label{MREG}
M(\Y; \Thet| Z, \eta, V_0) &=& \Delta H(\mathbf y; \Thet| Z, \eta, V_0)\nonumber
\\
& = & \frac{1}{2} \log|I_p + \eta^{-1} V_0 Z^\prime Z|
\\
& = & \frac{1}{2} \dsum_{j=1}^p \log ( 1+\eta^{-1} v_{0j} \lambda
_j ).\nonumber
\end{eqnarray}
From (\ref{MREG}) it is clear that the parameter (joint) information
is decreasing in $\eta$ and
increasing in $v_{0j}, \lambda_j$ and $\sigma^2_0$.
Thus, given the prior,
the information can be optimized through the choices of design
parameters $\lambda_j,  j=1,\ldots, p$, and
for given data (design), the information can be optimized through the
prior parameters $\sigma_0^2$ and
$v_{0j},  j=1,\ldots, p$.

The prior and posterior predictive distributions of a future outcome
$Y_\nu$ to be taken at a point $\z_\nu$
are normal $N (\z_\nu^\prime\bolds{\mu}_k,  \sigma^2_k \z_\nu
^\prime V_k \z_\nu+\sigma^2_1 ),  k=0,1$
and
\begin{eqnarray}
\label{MREGPRED}
&&M(\Y;Y_\nu|\z_\nu, Z, \eta, V_0)\nonumber
\\
 &&\quad= \Delta H(\mathbf y;Y_\nu|\z_\nu,
Z, \eta, V_0)
\\
&&\quad =
\frac{1}{2} \log \biggl( \frac{\eta^{-1}\z_\nu^\prime V_0 \z_\nu+
1}{\z_\nu^\prime V_1 \z_\nu+1} \biggr).\nonumber
\end{eqnarray}

Parts (a) and (b) of Theorem \ref{BDPI} give
$\Delta H[\mathbf y; (\Thet,Y_\nu)|\break\z_\nu, Z, \eta, V_0] = \Delta
H(\mathbf y; \Thet|Z, \eta, V_0)$
and $M[\Y; (\Thet,\break Y_\nu)| \z_\nu, Z, \eta, V_0] = M(\Y; \Thet|Z,
\eta, V_0)$.
Therefore all existing results for $M(\Y; \Thet|Z, \eta, V_0)$ apply
to the joint parameter and
predictive information, as well.
Part (c) of Theorem \ref{BDPI} provides an additional insight:
$M(\Y;Y_\nu|\z_\nu, Z, \eta, V_0) \leq M(\Y; \Thet|Z, \eta, V_0)$.
These relationships hold for multiple predictions, as well.

\subsection{Optimal Designs} \label{ANOVA}

Several authors have studied parameter information in the context of
experimental design.
It is clear from (\ref{MREG}) that given $V_0=I_p$ and the trace
$\mbox{Tr}(Z^\prime Z)= \sum_{j=1}^p\lambda_j$,
the optimal parameter information design is obtained when all
eigenvalues are equal,
$\lambda_j = \bar{\lambda}=\frac{1}{p}\sum_{k=1}^p\lambda_k$,
which gives the Bayesian D-optimal design (see Chaloner and Verdinelli,
\citeyear{1995Chaloner}, for references).
That is, with the uncorrelated prior the information optimal design is
orthogonal.
For the case of weak prior information, $\sigma^2_0 \to\infty$,
maximizing the expected parameter information gain is equivalent to the
classical criterion of D-optimality. If the experimental information
is weak, then the Bayesian criterion reduces to the classical criterion of
A-opti\-mality when $V_0=I_p$ (Polson, \citeyear{1993Polson}).
Verdinelli, {Polson} and {Singpurwalla} (\citeyear{1993Verdinelli}) used the predictive information optimal design
for accelerated life testing.

To illustrate implications of Theorem \ref{BDPI} for design we
consider the simple case
when $x_{ij}\in\{0,1\}$. This is a one-way ANOVA structure,
when the averages (parameters) as well as contrasts between the
individual outcomes are of interest.
In this case, $\operatorname{Tr}(\Lambda)=\sum_{j=1}^p n_j=n$ and the
design parameters are $\lambda_j= n_j$.
The following proposition gives the optimal designs according to the
parameter (joint) information
$M(\Y; \Thet)$ and predictive information $M(\Y; Y_\nu)$.

\begin{Proposition} \label{ALLOCATN}
Given $\eta, V_0$ and $\sum_{j=1}^p n_j=n$:
\begin{longlist}[(a)]
\item[(a)]
The optimal sample allocation scheme according to the parameter (joint)
information $M(\Y; \Thet)$ is
\begin{equation}
\label{OPTNIPARA}
\cases{
n^*_1 = \dfrac{n}{p} + \dfrac{\eta}{p}\displaystyle \sum_{j=2}^p ( v_{0j}^{-1}
- v_{01}^{-1} ),
\vspace*{2pt}\cr
n^*_j = n^*_1 - \eta ( v_{0j}^{-1} - v_{01}^{-1} ), \quad j=2,\ldots,p,
}
\end{equation}
and the minimum sample size is determined by
$n^*_1 > \max \{ ( v_{0j}^{-1} - v_{01}^{-1} )\eta,  j=2,\ldots,p \}$.
\item[(b)]
The information optimal sample allocation sche\-me according to the
predictive information $M(\Y; Y_\nu)$
for prediction at $\z_\nu$ is
\begin{equation}
\label{OPTNIPRED}
 \hspace*{5pt}\quad \cases{
n^*_1 = \dfrac{|z_{\nu1}|n}{\sum_{j=1}^p |z_{\nu j}|}
\cr
{}\qquad + \dfrac{\eta}{\sum_{j=1}^p |z_{\nu j}|} \displaystyle\sum_{j=2}^p (
v_{0j}^{-1} - v_{01}^{-1} ),
\vspace*{2pt}\cr
n^*_j = \dfrac{|z_{\nu j}|}{|z_{\nu1}|} n^*_1
- \dfrac{\eta}{|z_{\nu1}|} ( |z_{\nu1}| v_{0j}^{-1} - |z_{\nu j}|
v_{01}^{-1} ),  \vspace*{2pt}\cr\quad j=2,\ldots,p,
}\hspace*{-10pt}
\end{equation}
and the minimum sample size is determined by
$n^*_1 > \max
\{ \frac{\eta}{|z_{\nu j}|} ( |z_{\nu1}| v_{0j}^{-1} - |z_{\nu j}|
v_{01}^{-1} ),  j=2,\ldots,p \}$.
\end{longlist}
\end{Proposition}

\begin{pf}
See the \hyperref[app]{Appendix}.
\end{pf}

Note that by Theorem \ref{BDPI}, the maximum predictive information
attained with
optimal design (\ref{OPTNIPRED}) is dominated by the parameter information:
\begin{eqnarray*}
M(\Y; \Thet | n_i^*, \z_\nu)&=&M[\Y; (\Thet, Y_\nu)| n_i^*, \z
_\nu]
\\
&\geq& M(\Y;Y_\nu| n_i^*, \z_\nu).
\end{eqnarray*}

\begin{figure}
\tabcolsep=0pt
\begin{tabular}{c}

\includegraphics{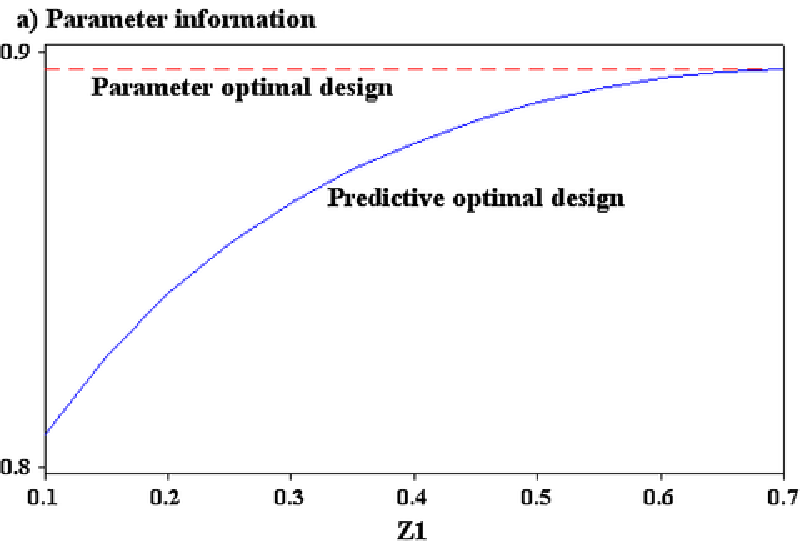}
\\[6pt]

\includegraphics{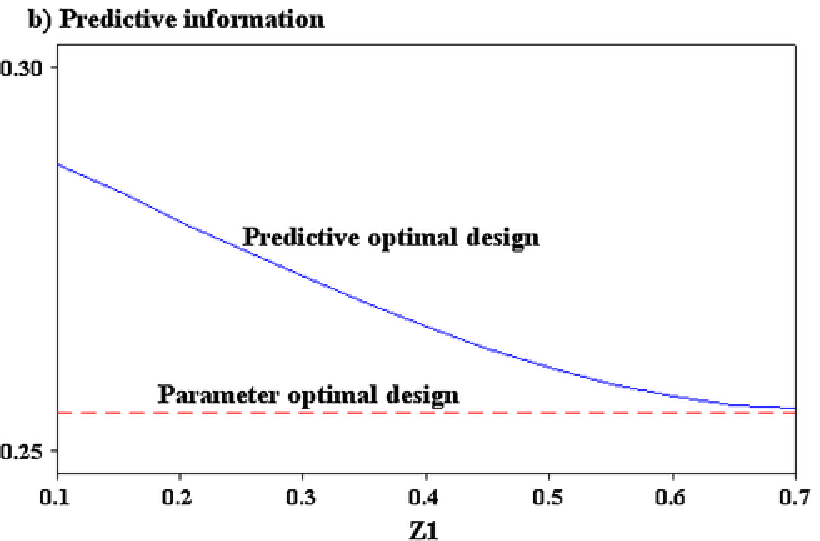}

\end{tabular}
\caption{Parameter information per dimension $M(\Y; \Thet|
Z, \eta)/p$ and
predictive information $M(\Y; Y_\nu| Z, \z, \eta)$ under the predictive
and parameter optimal designs against $z_{\nu1},  \z_\nu^\prime\z
_\nu=1$
for $p=2,  n=10, \eta=1, v_{01}=v_{02}=1$.}\label{fig2}
\end{figure}

\begin{Example}
Let $p=2, n=10, v_{01}=v_{02}=1, \eta=1$ and $\z_\nu^\prime\z_\nu=1$.
Figure \ref{fig2}(a) shows the plots of the parameter information
measures under the parameter and predictive optimal designs against
$z_{\nu1}$. In order to make the two information measures
dimensionally comparable, we have plotted information per parameter
$\bar{M}(\Y; \Theta | n_i^*)=M(\Y; \Theta | n_i^*)/p$.
Figure~\ref{fig2}(b) shows the plots of the predictive information
measures under the parameter and predictive optimal designs.
Note that the vertical axes of the two panels are different.
These plots show that the parameter (joint) information per dimension is much
higher than the predictive
information even when the design is optimal for prediction and not for
the parameter.
The dashed lines show the information quantities for the D-optimal
design, which is optimal for
the parameter (joint) and for prediction at the diagonal
$z_1=z_2=1/\sqrt{2}\approx0.707$.
The sample is least informative for prediction in this direction.
We note that the loss of information for prediction
is not nearly as severe as the loss of information about the parameter.
This is due to the fact that by Theorem \ref{BDPI}, the parameter
information measures the joint information
about the parameter and prediction and is inclusive of the predictive information.
Thus, use of the D-optimal design
would be preferable if the experimenter has interest in inference about
the parameter as well as about a prediction.
\end{Example}

\subsection{Optimal Prior Variance} \label{REG}

Next we illustrate application to developing prior in the context of a
Bayesian solution to the collinearity problem.
When the regression matrix $X$ is ill-conditioned, posterior inference
about individual
parameters is unreliable. The effects of collinearity on the posterior
distribution
and compensating for the collinearity effects by using $V_0=I_p$ were
discussed by Soofi (\citeyear{1990Soofi}).
In the orthogonal prior variance case $\sum_{j=1}^p v_{0j} =p$ is
distributed uniformly
among the components of $V_0$.
The following proposition gives an optimal prior variance
allocation according to the parameter (joint) information $M(\Y; \Thet
)$ that will be useful
when $X^\prime X$ is nearly singular.

\begin{Proposition}  \label{ALLOCATV}
Let $\lambda_1 \geq\cdots\geq\lambda_p,  \sum_{j=1}^p \lambda
_j = p$, and given $\eta$ and $\sum_{j=1}^p v_{0j} = c$.
The optimal prior variance allocation according to the parameter
(joint) information $M(\Y; \Thet)$ is
\begin{equation}
\label{OPTVIPARA}
\cases{
v_{01}^* = \dfrac{c}{p} + \dfrac{\eta}{p} \displaystyle\sum_{j=2}^p (\lambda
_j^{-1} - \lambda_1^{-1} ),
\cr
v_{0j}^* = v_{01}^* - \eta (\lambda_j^{-1} - \lambda_1^{-1} ),
\quad
j=2,\ldots,p,
}
\end{equation}
and the minimum prior variance is determined by
$v_{01}^* > (\lambda_p^{-1} - \lambda_1^{-1} )\eta$.
\end{Proposition}

\begin{pf}
See the \hyperref[app]{Appendix}.
\end{pf}

The optimal information prior (\ref{OPTVIPARA}) allocates prior
variances to the
components $\theta_j, j=1,\ldots, p$, based on the eigenvalues
$\lambda_1 \geq\cdots\geq\lambda_p$
of $X^\prime X$. So it is in the same spirit as Zellner's $g$ prior
(Zellner, \citeyear{1986Zellner}) where
$v_{0j} \propto\lambda_j^{-1}, j=1,\ldots, p$. In the same spirit,
West (\citeyear{2003West}) and Maruyama and George (\citeyear{2010Maruyama}) have defined generalized $g$
priors that
are applicable when $X$ is singular.
Our information optimal allocation scheme is another generalization of
the $g$ prior
tailored for the collinearity problem where $X$ is full-rank, but
nearly singular.

The optimal allocation scheme (\ref{OPTVIPARA}) can be represented in
terms of the condition indices
$\kappa_j = \sqrt{ \lambda_1/\lambda_j}$, $j=1,\ldots, p$, of
$X^\prime X$ as
\[
\cases{
\dfrac{\lambda_1 v^*_{01} + \eta}{\lambda_j v^*_{0j} + \eta}
= \kappa^2_j , &$ j=2, \ldots, p $,
\cr
\displaystyle\sum_{j=1}^p v^*_{0j} = c.
}
\]
The smallest portion of the total prior variance $v^*_{0p}$ is allocated
to the
component $\theta_p$ that corresponds to the smallest eigenvalue
$\lambda_p$
such that
$ \frac{\lambda_1 v^*_{01} + \eta}{\lambda_p v^*_{0p} + \eta} =
\kappa^2 $,
where $\kappa = \kappa(X^\prime X) = \sqrt{ \lambda_1/\lambda_p}$
is the \textit{condition number} of $X^\prime X$
which is used for collinearity diagnostics (Stewart, \citeyear{1987Stewart}; Soofi, \citeyear{1990Soofi};
Belsley, \citeyear{1991Belsley}).

In some prediction problems, the prediction point $\z_\nu$ is given.
For example, in the accelerated life testing, $\z_\nu$ is the environmental
condition and the experiment must be designed such that prediction at
$\z_\nu$ is
optimal. The information decomposition (\ref{MREG}) provides the clue when
the quantity of interest is the mean response $Q=E(Y|\z_\nu)$.
The components of $\bolds{\theta}=(\theta_1, \ldots, \theta_p)^\prime$
are independent,
a priori and a posteriori, and from (\ref{MREG}),
$M(\theta_j, \Y| Z, \eta, V_0)=0.5\log ( 1+\eta^{-1} v_{0j}
\lambda_j )$.
Under the orthogonal prior, the sample is most informative about
the linear combination of the regression coefficients $\theta_1 = \mathbf
g_1^\prime\bet$
where $g_1$ is the first eigenvector of $X^\prime X$.
Thus the optimal design for the expected response at a covariate vector
$\z_\nu$
is $X^*$ such that $\z_\nu$ is the first eigenvector of $I_p + \eta
^{-1} V_0 X^{*\prime}X^*$.
Under the uncorrelated prior or weak prior, $X^*$ is frequentist \textit{E-optimal}
design, which can be different than the designs that are optimal with
respect to parameter (joint) information.
The optimal allocation scheme (\ref{OPTVIPARA}) provides improvement
to the orthogonal prior
for prediction of the expected response
when $\z_\nu$ is in the space of the eigenvectors corresponding to
the large eigenvalues.

\begin{figure}
\tabcolsep=0pt
\begin{tabular}{c}

\includegraphics{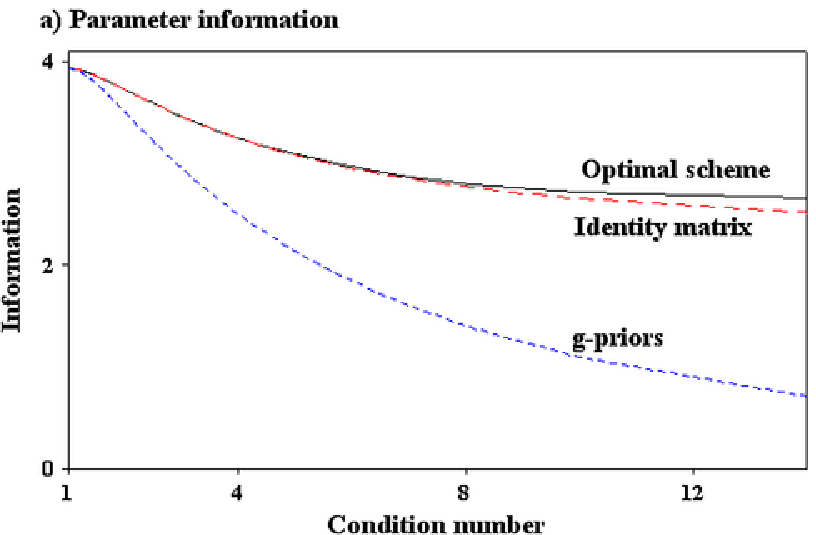}
\\[6pt]

\includegraphics{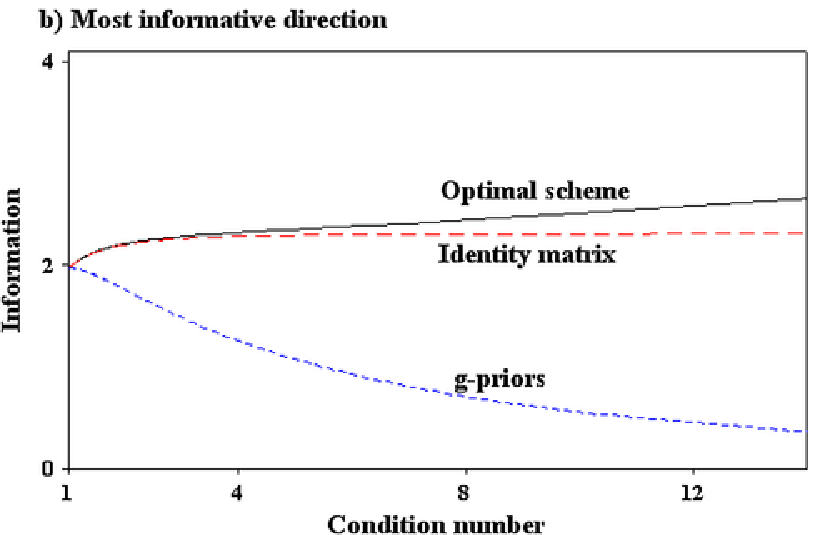}

\end{tabular}
\caption{Parameter information $M(\Y; \Thet| Z, \eta)$ and
information
for the most informative direction for prediction
of the expected response $M(\Y; \theta_1 | Z, \eta)$ for three types
of prior
variance allocations ($p=2,  c=100, \eta=1$).}\label{fig3}
\end{figure}

\begin{Example}
Let $p=2,  c=100$ and $\eta=1$. Figure~\ref{fig3} compares information measures
for the optimal scheme, the orthogonal prior and $V_0 \propto\Lambda^{-1}$
which is used in some priors such as the $g$-prior.\break
Figure \ref{fig3}(a) shows the plots of parameter information $M(\Y; \Thet| Z,
\eta)$ against the condition number
$\kappa= \sqrt{ \lambda_1/\lambda_2}$ of $X^\prime X$.
Under all three priors, the parameter information $M(\Y; \Thet| Z,
\eta)$ decreases with $\kappa$,
that is, as the regression matrix descends toward singularity.
The parameter information under the optimal scheme slightly dominates
the measure under the orthogonal prior, and both dominate the
information under the $g$-prior which
deteriorates quickly with collinearity.
By Theorem \ref{BDPI}, the parameter information measure is the joint
information
about the parameter and prediction and is inclusive of the predictive
information.
Figure~\ref{fig3}(b) shows $M(\Y; \theta_1 | Z, \eta)$ for the direction of
the first eigenvector
$\theta_1=G_1^\prime\bet$, that is,
the most informative direction for prediction of the expected response.
The optimal and orthogonal priors improve the information under collinearity,
but the measure for the $g$-prior deteriorates quickly.
\end{Example}

\section{Exponential Family} \label{EXPO}

Consider distributions in the exponential family that provide
likelihood functions in the form of
\begin{equation}
\label{EXPF}
\mathcal{L} (\theta) \propto \theta^n e^{-\theta s_n}, \quad \theta>0,
\end{equation}
where $s_n$ is a sufficient statistic for $\theta$.
This is the likelihood function for an important class of models referred to as
 the time-transformed exponential (TTE) (Barlow and Hsiung, \citeyear{1983Barlow}).
The TTE models are usually defined in terms of the survival function
$\bar{F}(y|\theta)=\exp\{-\theta\phi(y)\},  y \geq0$,
where $\phi(y)=-\log\bar{F}_0$ and $\theta$ is the ``proportional
hazard.''
The density functions of the TTE models are in the form of
\begin{equation}
\label{TTE}
f (\phi(y)|\theta )=\theta\phi^\prime(y) e^{-\theta\phi(y)},
\end{equation}
where $\phi(y)$ is a one-to-one transformation of $Y$ with the exponential
distribution $f(y|\theta)=\theta e^{-\theta y}$. For TTE models
$s_n=\sum_{i=1}^n \phi(y_i)$.
Examples include the
exponential $\phi(y)= y,  y \geq0$, Weibull $\phi(y)= y^q,  y \geq
0$,
Pareto Type I $\phi(y)= \log(y/a),  y \geq a >0$, Pareto Type II
$\phi(y)= \log(1+y),  y \geq0$,
Pareto Type VI $\phi(y)= \log(1+y^a),  y \geq0 ,  a >0$ and the
extreme value $\phi(y)= e^y$.

The family of conjugate priors for (\ref{EXPF}) is gamma $\mathcal{
G}(\alpha, \beta)$ with density function
\begin{equation}
\label{PRIOR}
f(\theta) = \frac{\beta^\alpha}{\Gamma(\alpha)} \theta^{\alpha
-1} e^{-\beta\theta}.
\end{equation}
The posterior distribution is $\mathcal{G}(\alpha+ n, \beta+ s_n)$.

The information in the observed sample is given by
\[
\Delta H(\mathbf y; \Theta)= H_\mathcal{G} (\alpha) - H_\mathcal{G}(\alpha+ n)
+ \log \biggl(1+\frac{s_n}{\beta} \biggr),
\]
where $H_\mathcal{G} (\alpha)$ is the entropy of $\mathcal{G}(\alpha, 1)$
given by
\[
H_\mathcal{G} (\alpha) = \log\Gamma(\alpha) -( \alpha-1) \psi(\alpha
) +\alpha,
\]
and $\psi(\alpha)=\frac{d \log\Gamma(\alpha)}{d \alpha}$ is the
digamma function.

For the TTE family (\ref{TTE}), the marginal distribution of $s_n$ is
inverted beta
(beta prime) distribution with density
\[
f(s_n) = \frac{1/\beta}{B(\alpha,n)}
\frac{(s_n/\beta)^{n-1}}{ (1+ s_n/\beta)^{\alpha+ n}},  \quad  s_n
\geq0,
\]
where $B(\alpha,n)$ is the beta function.
Using $E_{s_n} \{\log (1+\frac{s_n}{\beta} ) \} =\psi(\alpha
+n)-\psi(\alpha)$,
the expected information for all models with likelihood functions in
the form of (\ref{EXPF}) is
\begin{eqnarray}
\label{MITHETA}
M[\Y; (\Theta; Y_\nu)] &=& M(\Y; \Theta)\nonumber
\\
&=& H_\mathcal{G} (\alpha) - H_\mathcal{G}(\alpha+ n)
\\
&&{}+ \psi(\alpha+ n) -
\psi(\alpha).\nonumber
\end{eqnarray}

An interesting property of (\ref{MITHETA}) is the following recursion:
\begin{eqnarray}
\label{MMKG}
M(\mathbf{Y}_n;\Theta| \alpha) &=& M(\mathbf{Y}_{n-1};\Theta| \alpha)\nonumber
\\[-8pt]\\[-8pt]
&&{}+
K_\mathcal{G}(\alpha+ n-1),\nonumber
\end{eqnarray}
where $\mathbf{Y}_n$ and $\mathbf{Y}_{n-1}$ are vectors of dimensions $n$ and
$n-1$, and
\begin{equation}
\label{KGV}
K_\mathcal{G}(v)=K(\mathcal{G}_v\dvtx \mathcal{G}_{v+1}) = \frac{1}{v} + \psi(v )
- \log v
\end{equation}
is the Kullback--Leibler information between $\mathcal{G}_v=\mathcal{G}(\nu,
\beta)$
and $\mathcal{G}_{v+1}=\mathcal{G}(\nu+1, \beta)$. The recursion (\ref{MMKG})
is found using $\psi(\alpha+ 1) = \psi(\alpha) + \frac{1}{\alpha}$.
By (\ref{MMMM}), $M(\mathbf{Y}_n;\Theta|\break\mathbf{Y}_{n-1})= K_\mathcal{
G}(\alpha+ n-1)$.
That is, on average, the incremental contribution of an additional observation
is equivalent to the information divergence due to one unit increase of
the prior shape parameter.

The prior predictive distribution for the exponential model is Pareto
$\mathcal{P}(\alpha, \beta)$ with density function
\[
f(y_\nu) = \frac{\alpha\beta^\alpha}{(\beta+ y_\nu)^{\alpha
+1}},  \quad  y_\nu\geq0 .
\]
The posterior predictive distribution $f(y_\nu| \mathbf y)$ is also Pareto
with the updated parameters $\mathcal{P}(\alpha+ n, \beta+ s_n)$. The
predictive information measures are given by
\begin{eqnarray}\label{MIPRED}
\Delta H(\mathbf y; Y_\nu) & = & H_\mathcal{P}(\alpha)-H_\mathcal{P}(\alpha
+n)\nonumber
\\[-1pt]
&&{}-\log \biggl(1+\frac{s_n}{\beta} \biggr),\nonumber
\\[-1pt]
M(\Y; Y_\nu) & = & H_\mathcal{P}(\alpha)-H_\mathcal{P}(\alpha+n)\nonumber
\\[-8pt]\\[-8pt]
&&{}- \psi
(\alpha+ n) + \psi(\alpha),\nonumber
\end{eqnarray}
where $H_\mathcal{P}(\alpha) = \frac{1}{\alpha} - \log\alpha+ 1$ is
the entropy of $\mathcal{P}(\alpha, 1)$.

By invariance of the mutual information, the expected predictive
information for TTE family (\ref{TTE}) is given by  (\ref{MIPRED}).

By Theorem \ref{BDPI}, $\Delta H [s_n;(\Theta, Y_\nu)]=\Delta H(\mathbf
y; \Theta)$,\break
$M[\Y; (\Theta, Y_\nu)]=M(\Y; \Theta)$ and $M(\Y; Y_\nu)\leq
M(\Y;\break \Theta)$.
The following theorem gives a more specific pattern of relationships.

\begin{Theorem}  \label{TTGF}
The following results hold for the TTE family (\ref{TTE}) and gamma
prior (\ref{PRIOR}):
\begin{longlist}[(a)]
\item[(a)]
$M(\Y; \Theta| \alpha)$ and $M(\Y; Y_\nu| \alpha)$ are
decreasing functions of $\alpha$, increasing functions of $n$ and as
$n \to\infty$,
$M(\mathbf{Y}_{n+1}; \Theta| \alpha) - M(\mathbf{Y}_n; \Theta| \alpha)
\to0$ and
$M(\Y;Y_\nu| \alpha) \to K_\mathcal{G}(\alpha)$.
\item[(b)]
$M(\Y; \Theta| \alpha) = M(\Y;Y_\nu| \alpha) + M(\Y; \Theta|
\alpha+1)$,
where $M(\Y; \Theta| \alpha+1)$ is the sample information with gamma prior
$\mathcal{G}(\alpha+1, \beta)$.
\item[(c)]
$M(\Y; \Theta| \alpha) -M(\Y;Y_\nu| \alpha)$ increases with
$\alpha$ and with $n$.
\end{longlist}
\end{Theorem}

\begin{pf}
For (a), it is known that the expected parameter and predictive
measures are increasing functions of $n$.
It was shown by Ebrahimi and Soofi (\citeyear{1990Ebrahimi}) that for the exponential model,
$M(\Y; \Theta| \alpha)$ is decreasing in $\alpha$. By the
invariance of
the mutual information the same result holds for the TTE family.
The limits are found by noting that $K_\mathcal{G}(v) \to0$ as $v \to
\infty$.
The expected predictive measure decreasing in $\alpha$ is found by
taking the derivative,
using series expansion\vadjust{\goodbreak} of the trigamma function
$\psi^\prime(u) =\sum_{k=1}^\infty\frac{1}{(u +k)^2}$
(Abramowitz and Stegun, \citeyear{1970Abramowitz}),
and an induction on $n$ that shows the derivative is negative.
Part (b) is found using recursion $\psi(\alpha+ 1) = \psi(\alpha) +
\frac{1}{\alpha}$.
Part (c) is implied by (a) and (b). The difference is $M(\Y; \Theta|
\alpha+1)$ which is
increasing (decreasing) in $n$ ($\alpha$).
\end{pf}

By part (a) of Theorem \ref{TTGF}, the parameter and predictive
information both increase with $n$.
Part (b) of Theorem \ref{TTGF} gives the relationship between
the parameter (joint) and the predictive information measures. Part (c)
indicates that under conditional
independence, the parameter (joint) information grows faster than the
predictive information
with the sample size.

\begin{figure} \label{AA}
\tabcolsep=0pt
\begin{tabular}{c}

\includegraphics{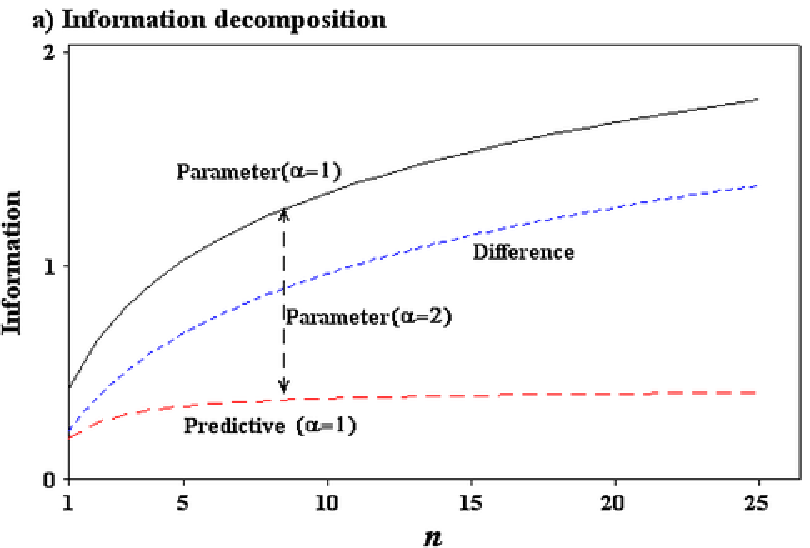}
\\[6pt]

\includegraphics{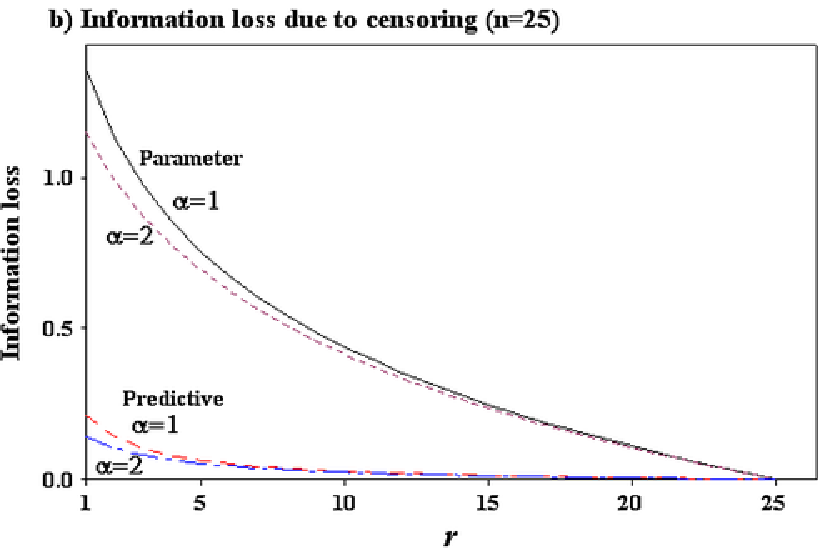}

\end{tabular}
\caption{Decomposition of the joint (parameter) information
$M(T_r; \Theta|\alpha, n)$
into predictive information $M(T_r; Y_\nu|\alpha, n)$ and $M(T_r;
\Theta|\alpha+1, n)$
and loss of information $M(T_n; \Theta|\alpha)-M(T_r; \Theta|\alpha
)$ due to Type II censoring
of exponential data.}\label{fig4}
\end{figure}

\begin{Example}
As an application, consider Type II censoring where observing the number
of failures is a design parameter.
For the exponential model, the sufficient statistic for $\theta$ in
(\ref{EXPF}) is
the total time under the test
\[
t_r = y_1 + \cdots+ y_{r-1} + (n-r+1) y_r,  \quad  r \leq n,
\]
where $y_1 \leq y_2 \leq\cdots\leq y_n$ are the order statistics of
a sample of size $n$.
The parameter information $M(T_r;\break \Theta| n)$ is
given by (\ref{MITHETA}) and the predictive information $M(T_r; Y_\nu
|\alpha, n)$
is given by (\ref{MIPRED}) with $n=r$. Ebrahimi and Soofi (\citeyear{1990Ebrahimi})
examined the loss of
information about the exponential failure rate.
By part (a) of Theorem \ref{TTGF}, censoring also results in loss of
predictive information.
As in the case of parameter information, the loss of predictive information
can be compensated by the prior parameter $\alpha$.
Figure~\ref{fig4} shows plots of the expected parameter and predictive
information measures.
Figure~\ref{fig4}(a) illustrates the information decomposition part [Theorem
\ref{TTGF}, part (b)]
for $\alpha=1$ as function of $n$. The parameter information and
predictive information are both
increasing in $n$. The parameter information increases at a faster rate
than the predictive
information. In this case, the difference between the parameter
and predictive information is $M(\mathbf y; \Theta| \alpha+1)$, also
shown in Figure~\ref{fig4}(a).
These information measures are decreasing in $\alpha$.
Figure~\ref{fig4}(b) shows the plots of loss of information due to Type II censoring
for $n=25$ and $\alpha=1,2$. We note that the predictive information
loss is not as severe as the
parameter information loss. As seen in the figure, the information
losses can be recovered by
increase in prior precision.
\end{Example}

By part (a) of Theorem \ref{TTGF}, $M(\Y; \Theta)$ and $M(\Y; Y_\nu)$
are maximized by choosing $\alpha$ as small as possible.
It is natural to expect that the limiting case, which is the Jeffreys
prior $f(\theta) \propto\theta^{-1}$,
be optimal with respect to both the parameter and prediction
information. But its use is consequential.
Since the Jeffreys prior is improper, the expected parameter information
is given by the negative conditional entropy of the
posterior distribution, which is proper. However, unlike the mutual
information, the entropy is not
invariant under one-to-one transformations and the result depends on
the parametric function of
interest. For example, for the exponential model, the posterior
distribution of failure rate
$\theta$ is gamma $f(\theta|s_n)=\mathcal{G}(n, s_n)$ and its entropy is
$H[f(\theta|s_n)] = H_\mathcal{G}(n) - \log s_n$.
The distribution of $S_n$ is Pareto\break $f(s_n)\propto s_n^{-n}$ which is
proper for $n>1$
and $s_n \geq s_0>0$.
The expected parameter information, $\mathcal{I}(\Theta|S_n)=-\mathcal{
H}[f(\theta|s_n)]$,
is a decreasing function of $n$. But the posterior distribution of the
mean parameter $\mu=\theta^{-1}$
is inverse-gamma and information about the mean is increasing in $n$.
With the Jeffreys prior, the prior predictive distribution is also
improper. The posterior predictive
is Pareto $\mathcal{P}(n, s_n)$ and its entropy is
$H[f(Y_\nu|s_n)] = H_\mathcal{P}(n) + \log s_n$.
The expected predictive information is $\mathcal{I}(Y_\nu|S_n)=-\mathcal{
H}[f(Y_\nu|s_n)]$, which
is an increasing function of $n$.\vspace*{-2pt}

\section{Dependent Sequences} \label{DEPEND}

When the sequence of random variables $Y_i|\theta,  i=1,2, \ldots,$
is not conditionally independent,
the information provided by the sample about the parameter and
prediction jointly decomposes as\vspace*{-1pt}
\begin{eqnarray}
\label{MMM3}
\quad M[\Y; (\Theta, Y_\nu)] &= &M(\Y; \Theta) +M(\Y; Y_\nu|\Theta)\nonumber
\\[-6pt]\\[-6pt]
&= &M(\Y; Y_\nu) +M(\Y;\Theta| Y_\nu),\nonumber\vspace*{-2pt}
\end{eqnarray}
where $M(\Y, Y_\nu|\Theta) \geq0$ is the measure of conditional
dependence, hence the inequality becomes
equality for the case of conditional independence and (\ref{MMM3})
gives (\ref{MMM1}). Thus, for the
conditionally dependent sequence, $M[\Y; (\Theta, Y_\nu)]$ exceeds
$M(\Y; \Theta)$ by the\break amount $M(\Y; Y_\nu|\Theta)> 0$.
Also from (\ref{MMM3}), we find that\vspace*{-2pt}
\begin{eqnarray*}
M(\Y; \Theta) &\leq& M(\Y;Y_\nu) \quad   \mbox{if and only if}
\\[-2pt]
M(\Y; \Theta|\mathbf y_\nu) &\leq& M(\Y; Y_\nu|\Theta).\vspace*{-2pt}
\end{eqnarray*}
For strongly conditional dependent sequence, the second inequality is
plausible and
the predictive information $M(\Y, Y_\nu)$ can dominate the parameter
information $M(\Y; \Theta)$.

\begin{table*}[t]
\tablewidth=370pt
\caption{Formulas for uncorrelated, intraclass and serial correlation
models}\label{tab1}
\begin{tabular*}{370pt}{@{\extracolsep{4in minus 4in}}lccc@{}}
\hline
 & \textbf{Uncorrelated (UC)} & \textbf{Intraclass (IC)} & \textbf{Serial correlation (SC)}\\
\hline
{Conditional sequence}
\\
\quad $|R | \theta|$ & 1 & $[1+(n-1)\rho](1-\rho)^{n-1}$
& $1-\rho^2$
\\
\quad $T_n(R|\theta)$  & $n$ & $\frac{n}{1+(n-1)\rho}$ &
$\frac{n-(n-2)\rho}{1+\rho}$
\\
\quad $\rho^2_{y_\nu, \mathbf y|\theta}$& 0 & $\frac{n\rho
^2}{1+ (n-1)\rho} $ & $\rho^2$
\\[3pt]
{Predictive sequence}
\\
\quad $\rho^2_p$ & $\frac{1}{1+\eta}$ & $\frac{1+\eta
\rho}{1+\eta}$ & $\frac{1+\eta\rho^k}{1+\eta}$
\\[4pt]
\quad $\rho^2_{y_\nu, \mathbf y}$  & $\frac{n}{(1+\eta
)(n+\eta)}$ & $\frac{n\rho^2_p}{1+ (n-1)\rho_p}$
& Immediate future $\rho^2_p$
\\
\hline
\end{tabular*}
\end{table*}

In this section we first examine the effects of correlation between
observations on the
information about the mean parameter and prediction where the data are
normally distributed.
We then consider order statistics where no particular prior distribution
and model for the likelihood function are assumed.\vspace*{-1pt}

\subsection{Intraclass and Serially Correlated Models} \label{UCICSC}

We consider the intercept linear model $f(\mathbf y|\theta)=N(\theta\z,
\sigma_1^2R)$,
where $\z$ is an $n \times1$ vector of ones and $R=R|\theta=[\rho
_{ij|\theta}]$ is a known correlation matrix.
By invariance of the mutual information, the results hold for all
distributions of variables
that are one-to-one transformations of elements of $\mathbf y$, for example,
log-normal model. As~before, $\sigma_1^2>0$ is known and $f(\theta)=N(\mu_0, \sigma_0^2)$.
The posterior variance is given by
$\sigma^2_{\theta|\mathbf y} =\sigma_0^2 [1+ T_n(R)\eta^{-1} ]^{-1}$,
where $T_n(R)=\z^\prime R^{-1}\z$ is the sum of all elements of $R^{-1}$.
The parameter information is given by\vspace*{-2pt}
\begin{equation}
\label{MNPARA}
M(\Y; \Theta|R) = 0.5\log \bigl(1 + \eta^{-1}T_n(R) \bigr).\vspace*{-2pt}
\end{equation}

The following representations facilitate computation and study of the
predictive and joint information
measures. If $Y_\nu$ and $Y_\nu|\mathbf y$ are normal, then the
predictive information is given by
\begin{eqnarray}
\label{MNPRED}
 M(\Y;Y_\nu)&=& - 0.5\log (1- \rho^2_{y_\nu, \mathbf y} )\nonumber
\\[-8pt]\\[-8pt]
  &=& 0.5\log
[C^{-1}]_{\nu\nu},\nonumber
\end{eqnarray}
where $\rho^2_{y_\nu, \mathbf y}$ is the square of unconditional multiple
correlation coefficient
of the regression of $Y_\nu$ on $\mathbf y$, $C=[c_{ij}],  i,j=1,\ldots
,n+1$, denotes the correlation
matrix of the $(n+1)$-dimensional vector $(\Y, Y_\nu)$, and
$[C^{-1}]_{\nu\nu}$ denotes the
$(\nu,\nu)$ element of $C^{-1}$.

The joint information about the parameter and prediction can be
computed by the first decomposition in~(\ref{MMM3}),
\begin{equation}
\label{MNJOINT}
\qquad M[\Y; (\Theta, Y_\nu)] =M(\Y; \Theta|R) +M(\Y, Y_\nu|\Theta),
\end{equation}
where $M(\Y; \Theta|R)$ is given in (\ref{MNPARA}) and
the measure of conditional dependence can be computed similarly to~(\ref{MNPRED}):
\begin{eqnarray}
\label{MNSAMPLE}
M(\Y; Y_\nu|\Theta)&=&-0.5\log (1-\rho^2_{y_\nu, \mathbf y|\theta} )\nonumber
\\[-8pt]\\[-8pt]
&=& 0.5\log[C^{-1}|\theta]_{\nu\nu} \geq0,\nonumber
\end{eqnarray}
where $\rho^2_{y_\nu, \mathbf y|\theta}$ is the square of conditional
multiple correlation coefficient and
$ C|\theta=[c_{ij}|\theta],  i,j=1,\ldots,\break n+1$, is the correlation
matrix of conditional distribution
of $(\Y, Y_\nu)$, given $\theta$.
Note that $C|\theta$ includes $R$ and an additional row and column for
$Y_\nu$.

Measures such as the determinant $|R|$ and condition number
$\kappa(R)= \sqrt{\lambda_1/\lambda_n}$, where $\lambda_1< \cdots
< \lambda_n$ are eigenvalues
of $R$, can be used to rank dependence of the normal samples.
However, in general, these measures do not provide a unique ranking.
In order to rank the dependence uniquely as well as for ranking the
predictive information in terms of sample
dependence, we assume some structures for $R$.
We consider two important models: the intraclass (IC) model with
$\rho_{ij|\theta}=\rho$ for all $i \neq j$, and the serial
correlation (SC) model with
$\rho_{i,i\pm k|\theta}=\rho^k \geq0,  k>0$. Dependence within
each of these models
and between the two models is ranked uniquely by $|R|$ and $\kappa(R)$.

Table~\ref{tab1} shows $|R|$ and $T_n(R)$ for the IC, SC models along with the
independent (uncorrelated) model (UC).
The determinants and inverses of the IC and SC matrices are well known.
Using $T_n(R)$ in (\ref{MNPARA})
gives the parameter information.
The third row of Table~\ref{tab1} shows $\rho^2_{y_\nu, \mathbf y|\theta}$ which
is computed using (\ref{MNSAMPLE})
with $(n+1)$-dimensional IC and SC structures for $C$.
Table~\ref{tab1} also shows the square of unconditional (predictive) correlation
$\rho^2_p=c_{ij}$, which is
used in (\ref{MNPRED}) for computing the predictive information measures.
Computation of $\rho^2_p=c_{ij}$ is shown in the \hyperref[app]{Appendix}.
The last row of Table~\ref{tab1} shows the square of unconditional multiple
correlation coefficient
$\rho^2_{y_\nu, \mathbf y}$ computed from (\ref{MNPRED}). The predictive
measure for the SC model
is for the one-step prediction.

The effects of prior on the information quantities are induced through
$\eta$ which is
proportional to prior precision. Clearly (\ref{MNPARA}) is decreasing
in $\eta$. Using\vadjust{\goodbreak}
the last two rows of Table~\ref{tab1} it can be shown that
(\ref{MNPRED}) and the difference between (\ref{MNPARA}) and (\ref
{MNPRED}) are also decreasing in $\eta$.
Thus, the optimal prior for inference about the parameter and prediction
is to choose the prior variance as large as possible.

The following theorem summarizes the effects of the IC and SC
correlation structures on the normal
information measures (\ref{MNPARA})--(\ref{MNJOINT}).

\begin{Theorem}  \label{ICSC}
\begin{longlist}[(a)]
\item[(a)]
For all three models, $M(\Y; \Theta| \rho)$, $M(\Y; Y_\nu| \rho
)$ and $M[\Y; (\Theta, Y_\nu)| \rho)]$ increase with $n$
and decrease with $\eta$.
\item[(b)]
For both IC and SC models, $M(\Y; \Theta| \rho)$ decreases with
$\rho$, and
\[
M^{\mathit{IC}}(\Y; \Theta|\rho) \leq M^{\mathit{SC}}(\Y; \Theta|\rho) \leq
M^{\mathit{UC}}(\Y; \Theta),
\]
where the last equality holds if and only if $\rho=0$.
\item[(c)]
For both IC and SC models, $M(\Y; Y_\nu| \rho)$ increases with $\rho
$, and
\[
M^{\mathit{IC}}(\Y;Y_\nu|\rho) \geq M^{\mathit{SC}}(\Y; Y_\nu|\rho) \geq M^{\mathit{UC}}(\Y
; Y_\nu),
\]
where the last equality holds if and only if $\rho=0$.
\item[(d)]
For both IC and SC models, $M[\Y; (\Theta, Y_\nu)| \rho)]$
decreases in $\rho$ for $\rho\leq\rho_0(n, \eta)$
and increases in $\rho$ for $\rho> \rho_0(n, \eta)$, where $\rho
_0^{\mathit{IC}}(n, \eta)$ and $\rho_0^{\mathit{SC}}(n, \eta)$
are roots of quadratic equations and both are increasing in $n$ and
decreasing in $\eta$.
\end{longlist}
\end{Theorem}

\begin{pf}
(a) Can be easily seen by taking derivatives.
(b) It is also easy to see that for the correlated models $T_n(R)$ are
decreasing functions of $\rho$
and that $T^{\mathit{IC}}_n(R) \leq T^{\mathit{SC}}_n(R) \leq T^{\mathit{UC}}_n(R)=n$.
(c)\vspace*{1pt} This is implied by the facts that $\rho^{\mathit{IC}}_p >\rho^{\mathit{SC}}_p$ and
the predictive information increases with $\rho$, as expected.
(d)~Taking the derivative, $\rho_0(n, \eta)$ is given by the root of
$A_{n,\eta}\rho^2+B_{n,\eta}\rho+C_{n,\eta}=0$,
where $A_{n,\eta}^{\mathit{IC}}=n-1,  B_{n,\eta}^{\mathit{IC}}=2(1+n \eta^{-1}),
A_{n,\eta}^{\mathit{SC}}=1+(2n-1)\eta^{-1},  B_{n,\eta}^{\mathit{SC}}=1+(2n-1)\eta
^{-1}$ and $C_{n,\eta}=(1-n)\eta^{-1}$.
For each model there is only a unique positive solution.
\end{pf}

Theorem \ref{ICSC} formalizes the intuition that samples with stronger
dependence are
less informative about the parameter and more informative about prediction.
Since $M(\Y; \Theta|\rho)$ is increasing in $n$, one can compensate
the loss of parameter information due to the dependence by increasing
the sample size.
The following example illustrates these and some other noteworthy points.

\begin{Example}
\begin{longlist}[(a)]
\item[(a)]
Figure~\ref{fig5} shows plots of $M(\Y; \Theta|\rho)$ and $M(\Y;\break Y_\nu|
\rho)$ against sample size
for the UC model and the correlated models IC and SC with $\rho=0.50,0.75$.
Plots in panels (a) and (b) reveal the following features.
\begin{longlist}[(iii)]
\item[(i)]
All information measures are increasing in $n$.
\item[(ii)]
For the UC model, the parameter information is the highest
and has the fastest rate of increase with $n$, and the predictive
information is the lowest with the slowest
(almost flat) rate of increase.
\item[(iii)]
For the SC model, the parameter information is higher and increases
much faster than
the predictive information.
\item[(iv)]
For the IC model, the parameter
information is lower than the predictive information while both
measures have about the same
rates of increase.
\item[(v)]
Interestingly, for the UC and SC models, the differences between the
parameter and
predictive information measures grow with $n$ much faster than the
predictive information measures.
That is, the share of predictive information decreases with the sample size.
\item[(vi)]
As can be seen in Figure~\ref{fig5}(a), to gain about one unit (nit) of
information, we need
$n=3$ from the UC, and with $\rho=0.50,0.75$, we need $n=8,16$
observations under SC,
and $n=26, 37$ observations under IC models, respectively.
\end{longlist}

\begin{figure}
\tabcolsep=0pt
\begin{tabular}{c}

\includegraphics{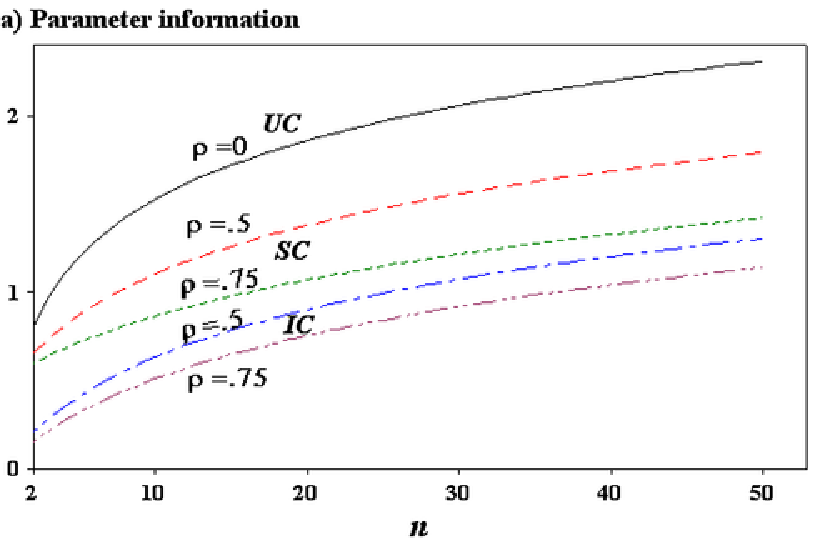}
\\[6pt]

\includegraphics{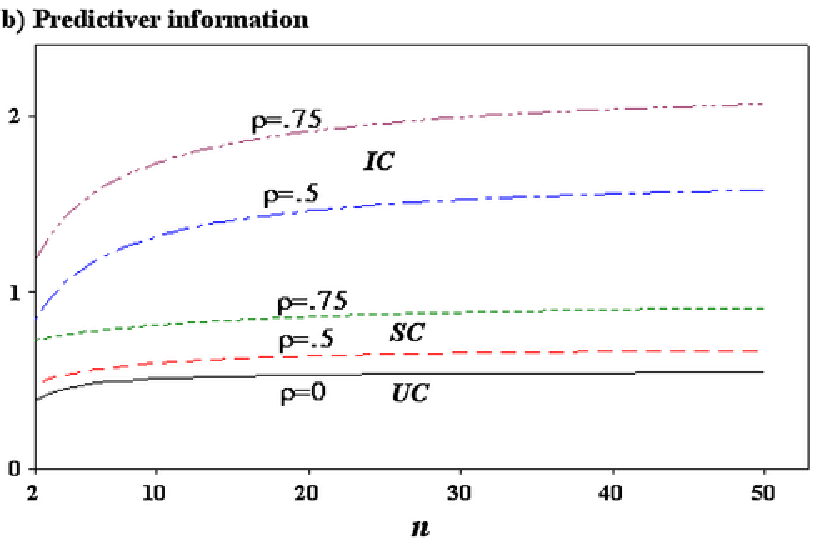}

\end{tabular}
\caption{The parameter information $M(\Y; \Theta| \rho)$
and predictive information
$M(\Y; Y_\nu| \rho)$ for the independent, IC and SC normal models
as functions of the sample size ($\eta=0.5$).}\label{fig5}\vspace*{-2pt}
\end{figure}

\begin{figure}
\tabcolsep=0pt
\begin{tabular}{c}

\includegraphics{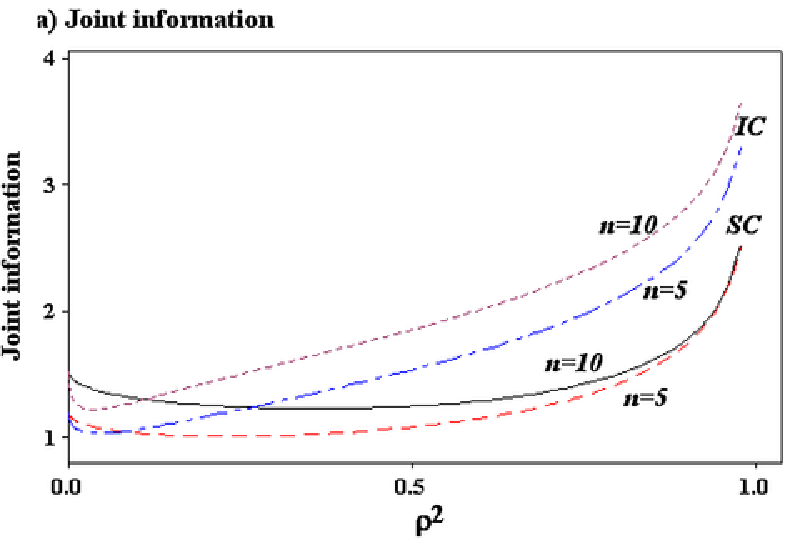}
\\[6pt]

\includegraphics{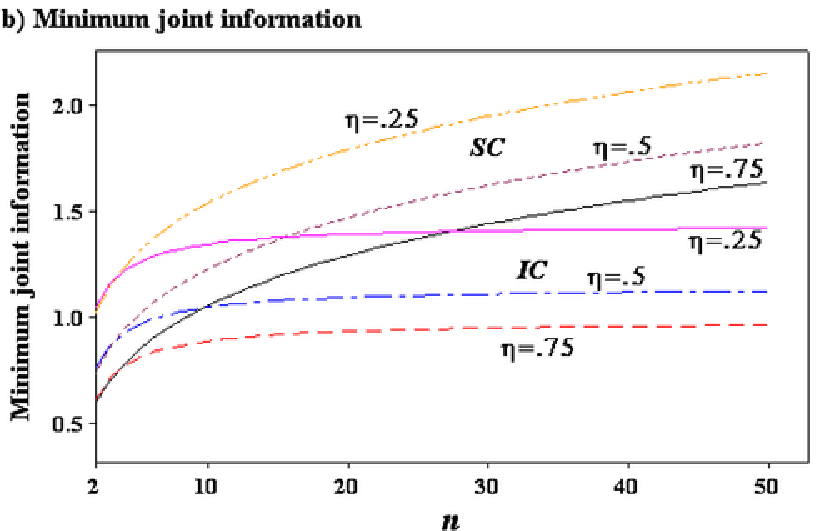}

\end{tabular}
\caption{The joint parameter and predictive information $M[\Y
; (\Theta, Y_\nu)| \rho]$ and
minima of the joint information $\min_{\rho} M[\Y; (\Theta, Y_\nu
)| \rho]$ for SC and IC normal models.}\label{fig6}
\end{figure}

\item[(b)]
Figure~\ref{fig6}(a) shows the plots of the joint information measures for the
SC and IC models as functions
of $\rho^2$ for $n=5,10$ and $\eta=0.5$. Note that the joint
information of the SC model
dominates the joint information of the IC model when dependence is
weak. After the minimum point,
the rate of growth of joint information for the IC model is steep and
the IC information measure
dominates the SC information measure when the dependence is rather strong.
\item[(c)]
Figure \ref{fig6}(b) shows the plots of the minimum joint information measures
for the SC and IC families as functions
of $n$ for $\eta=0.25, 0.5, 0.75$. These plots are useful for
determining sample size for each family
such that the minimum information exceeds a given value. For example,
to gain about 1.5 units (nits) of information from an SC sample with
unknown $\rho$, we need
$n=9, 25, 37$ with $\eta=0.25,0.50,0.75$, respectively. The plots show that
\[
M^{\mathit{IC}}_0[\Y; (\Theta, Y_\nu)|n, \eta] \leq M^{\mathit{SC}}_0[\Y; (\Theta,
Y_\nu)|n, \eta],
\]
where $M_0[\Y; (\Theta, Y_\nu)|n, \eta] = \min_{\rho} M[\Y;
(\Theta, Y_\nu)| \rho]$.\break
This inequality can be proved by substituting\break $\rho_0^{\mathit{IC}}(n, \eta)$
and $\rho_0^{\mathit{SC}}(n, \eta)$ in
the expressions for $T_n(R)$ and $\rho^2_{y_\nu, \mathbf y|\theta}$.
\end{longlist}
\end{Example}

\subsection{Order Statistics} \label{ORDSTAT}

\begin{figure}
\tabcolsep=0pt
\begin{tabular}{c}

\includegraphics{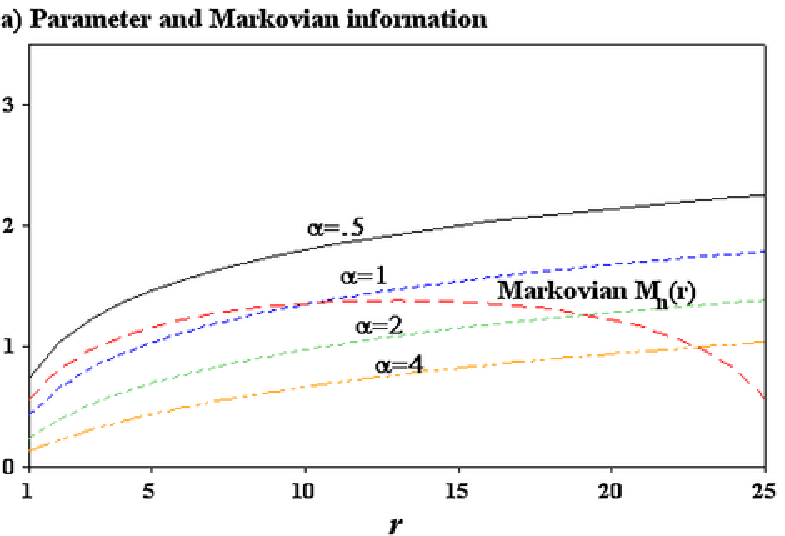}
\\
(a)\\[6pt]

\includegraphics{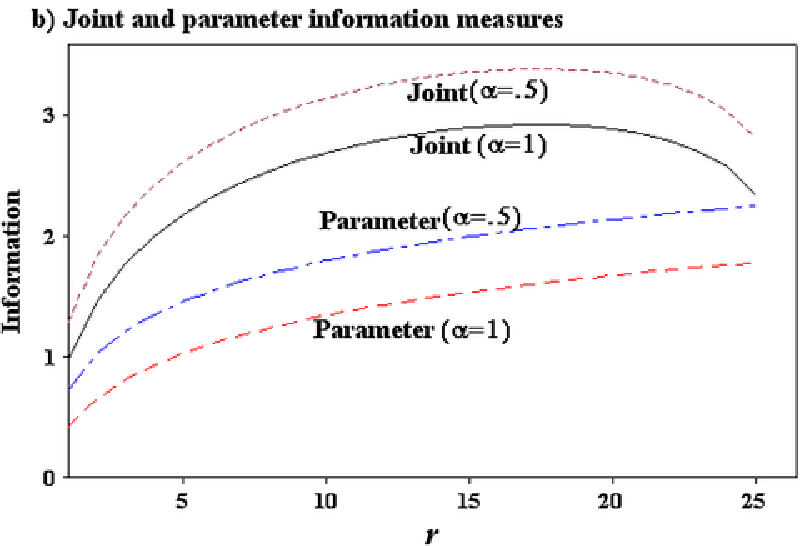}
\\
(b)
\end{tabular}
\caption{Expected information about the parameter $M(\Y_r,
\Theta)$
and the joint information about the parameter and prediction of the
$(r+1)$st order statistic
$M [\Y_r; (\Theta, Y_{r+1})]$ provided by the vector of preceding
order statistics $\Y_r$,
and the information due to the Markovian dependence between order
statistics ($n=26$).}\label{fig7}
\end{figure}

Let $Y_1 \leq Y_2 \leq\cdots\leq Y_n$ be the order statistics of
conditionally independent
sample $X_1, \ldots, X_n$ from a continuous distribution with density
function $g(x|\theta)$,
and let $\mathbf y_r=(y_1,\ldots,y_r),  r \leq n$. Conditional on $\theta
$, the order statistics
have a Markovian dependence structure (Arnold, \citeyear{1992Arnold}). The mutual
information between consecutive
order statistics is given by
\begin{eqnarray}
\label{MFRR1}
&&M (Y_r; Y_{r+1}|\theta)\nonumber
\\
&&\quad =  M_n(r)\nonumber
\\
&&\quad =  \log B(r+1,n-r+1) +\log(n+1) -1
\\
&&\qquad{}- r \{\psi(r)-\psi(n)\}\nonumber
\\
&&{}\qquad -(n-r)\{ \psi(n-r) -\psi(n)\};\nonumber
\end{eqnarray}
see the article by Ebrahimi, {Soofi} and {Zahedi} (\citeyear{2004Ebrahimi}).
That is, $M_n(r)$ is the measure of Markovian dependence between
order statistics of the independent sample conditional on $\theta$. It
was shown by
Ebrahimi, {Soofi} and {Zahedi} (\citeyear{2004Ebrahimi}) that
$M_n(r)$ is increasing in $n$, and for a given $n$, the information is
symmetric in $r$
and $n-r$, and attains its maximum at the median (see Figure~\ref{fig7}).
The next lemma gives generalizations of (\ref{MFRR1}).
All information functions are
conditional on $r$ and $n$, which will be suppressed when unnecessary.

\begin{Lemma}  \label{MOS}
Let $Y_1 \leq\cdots\leq Y_n$ denote the order statistics of random variables
$X_1 , \ldots, X_n $ which, given $\theta$, are independent and have
identical distribution
$g(x|\theta)$ and $\Y_r$ and $\Y_q$ denote the disjoint subvectors
of order statistics.
Then:
\begin{longlist}
\item[(a)]
$M (\Y_r;\Y_q|\theta)$ is free from the parent distribution
$g(x|\theta)$ and the prior
distribution $f(\theta)$.
\item[(b)]
For any two consecutive subvectors $\mathbf{Y}_r=\break (Y_{k+1}, \ldots,
Y_{k+r})$ and
$\Y_q=(Y_{k+r+1}, \ldots, Y_{k+r+q})$,\break $M (\Y_r; \Y_q|\theta) = M_n(k+r)$.
\end{longlist}
\end{Lemma}

\begin{pf}
Let $U= G(X)$. Then $U$ is uniform and its order statistics
$W_1 \leq W_2 \leq\cdots\leq W_n$ are given by $W_i = G(Y_i)$,
and $\W_r$ and $\W_q$
are the subvectors corresponding to $\Y_r$ and $\Y_q$.
Since $W_i= G(Y_i)$ is one-to-one, we have $M (\Y_r;\Y_q) =M (\W
_r;\W_q)$.
Furthermore the distribution of any subset of order statistics is
ordered Dirichlet
with parameters $n$ and the indices of the order statistics contained
in the subset,
hence $M (\Y_r;\Y_q) =M (\W_r;\W_q)$ is free from the parent
distribution $g(x|\theta)$.
Part (b) follows from $Y_1|\theta, \ldots, Y_n|\theta$ being a
Markovian sequence.
\end{pf}

It can easily be shown that
information provided by the first $r$ order statistics about the
parameter $M(\Y_r, \Theta)$
satisfies (\ref{MMMM}). The predictive distributions of order
statistics are given by
$f(y_i) = \int f(y_i|\theta)\* f(\theta) \,d \theta$,  $i=1,\ldots,n$.
Note that
$y_1 \leq y_2 \leq\cdots\leq y_n$ are the order statistics of a
sample of the
exchangeable sequence $X_1, \ldots, X_n$, unconditionally. The
following results provide
some insight about the parameter and predictive information for order
statistics.

\begin{Theorem}  \label{MSRTHETAY}
Let $M [\Y_r; (\Theta, Y_{r+1})]$ denote the information provided by
the first $r$ order
statistics about the parameter and for prediction of the next order
statistic jointly.
Then:
\begin{longlist}
\item[(a)]
$ M [(\Y_r; (\Theta, Y_{r+1})]= M(\Y_r;\Theta) + M_n(r) \geq\break M(\Y
_r;\Theta)$.
\item[(b)]
The following statements are equivalent:
\begin{longlist}[(ii)]
\item[(i)]
$M (\Y_r; Y_{r+1}) \geq (\leq)  M_n(r)$.
\item[(ii)]
$M(\Theta; Y_{r+1}) \geq(\leq)  M(\mathbf{Y}_{r+1};\Theta) - M(\Y_r;
\Theta) $,\break
where $\mathbf{Y}_{r+1}=(Y_1,\ldots, Y_r, Y_{r+1})$.
\end{longlist}
\end{longlist}
\end{Theorem}

\begin{pf}
Using the following decompositions of mutual information, we have
\[
M[(\Theta, Y_{r+1});\Y_r ] = M(\Y_r;\Theta) + M(\Y_r;
Y_{r+1}|\Theta).
\]
Applying part (b) of Lemma \ref{MOS} to the second term gives the
result (a).
For (b) we use the following decompositions of mutual information:
\begin{eqnarray*}
M [(\Y_r, \Theta); Y_{r+1}] & = & M(\Y_r; Y_{r+1}) + M(\Theta
;Y_{r+1}|\Y_r)
\\
&=& M(\Theta;Y_{r+1}) + M(\Y_r; Y_{r+1}|\Theta).
\end{eqnarray*}
Equating the two decompositions with $M(Y_{r+1};\Y_r|\break \Theta)=M_n(r)$
gives equivalence of (i) and
\begin{equation}
\label{MTHETASR}
M(\Theta; Y_{r+1}) \geq(\leq)  M(\Theta;Y_{r+1}|\Y_r).
\end{equation}
The equivalence with (ii) is obtained by solving
\[
M (\mathbf{Y}_{r+1}; \Theta) = M (\Y_r; \Theta) + M (\Theta;Y_{r+1}|\Y_r)
\]
for $M (\Theta;Y_{r+1}|\Y_r)$ and substituting in (\ref{MTHETASR}).
\end{pf}

Part (a) of Theorem \ref{MSRTHETAY} shows $M [\Y_r; (\Theta,
Y_{r+1})]$ is inclusive
of Lindley's measure reflecting the fact that conditional on $\theta$, order
statistics are dependent.
So the information provided by the first $r$ order statistics about
the parameter and for prediction of the next order statistic is more than
the information provided about the parameter. However, the excess
information amount
measures the Markovian dependence between order statistics of
the independent sample and does not depend on $g_{x|\theta}$ and
$f_\theta$.
An implication of this result is that
reference posterior corresponding to the prior that
maximizes the parameter information $M(\Y_r;\Theta)$ also remains optimal
with respect to $M [\Y_r; (\Theta, Y_{r+1})]$.

Part (b) of Theorem \ref{MSRTHETAY} gives the equivalence of the
orders of information in terms
of (i) the predictive and sample order statistics and (ii) the expected
information about
the parameter provided by an order statistic in terms of
the incremental amount of information provided about the parameter.

\begin{Example}\label{ROSEXO}
For the case of exponential model with the gamma prior,
the conditional distribution of $(r+1)$st order statistic given $\theta
$ and the
first $r$ order statistics is exponential with density
\begin{eqnarray}
&&f(y_{r+1} | y_r, \theta)\nonumber
\\[-8pt]\\[-8pt]
&&\quad= (n-r) \theta e^{- \theta(n-r) (y_{r+1} -
y_r)},   \quad y_{r+1} > y_r.\nonumber
\end{eqnarray}
The posterior predictive distribution of $(r+1)$st order statistic
given first $r$ order statistics
is Pareto with parameters $\alpha+r,  b_r=\frac{\beta+ t_r}{n-r}$ and
a location parameter $y_r$. Since entropy is location-invariant,\break
$H(Y_{r+1}| \mathbf y_r)$
is $H(Y_{r+1}| t_r, y_r, r, n) = H(Y_\nu| t_r, r) -\break \log(n-r)$.
Figure~\ref{fig7} illustrates some properties of these information measures for
the exponential model and $n=26$.
\begin{longlist}[(a)]
\item[(a)]
Figure \ref{fig7}(a)  shows plots of $M(\Y_r; \Theta|\alpha, n)=M(T_r; \Theta
|\alpha, n)$ for $\alpha=0.5, 1, 2, 4$,
superimposed by the Markovian dependence information measure $M_n(r)$
for the order statistics.
Since $M(T_r; \Theta|\alpha, n)$ is increasing in $r$, censoring
results in loss of information
about the parameter. Thus, without consideration of cost of the
experiment, $r^*=n=26$.
Since $M_n(r)$ is decreasing for $r$ larger than the median, censoring
beyond the median results in
gain of information about the next outcome.
\item[(b)]
Figure \ref{fig7}(b) shows the plots of the parameter information $M(\Y_r;
\Theta|\alpha, n)$
and joint information $M[\Y_r; (\Theta, Y_{r+1})|\alpha, r, n]$
computed using part (a) of Theorem \ref{MSRTHETAY} for $\alpha=0.5,
1$. We note that
$M[\Y_r; (\Theta,\break Y_{r+1})|\alpha, r, n]$ is not monotone because
the Markovian dependence information
measure $M_n(r)$\break decreases for the order statistics above the median.
The optimal $r$
for the joint parameter and predictive information, without
consideration of cost the experiment,
is $r^*=17 < n$. Thus, unlike the case of conditionally independent
model, the parameter information
utility and the joint parameter--predictive information utility lead to
different sampling plans.
\end{longlist}
\end{Example}

In Section \ref{EXPO} we noted that under the Jeffreys prior, at least
one observation is
needed for obtaining a proper posterior.
Following this idea more generally, we compare
the expected uncertainty change due to the first $r$ order statistics
with the first order statistic $r=1$, given by
\[
\mathcal{B}[\Y_r; (\Theta,Y_{r+1})] = H[(\Theta,Y_1)]- \mathcal{H}[(\Theta
,Y_{r+1})|\Y_r],
\]
where $\mathcal{H}[(\Theta,Y_{r+1})|\Y_r] = E_{s_r} \{H[(\Theta
,Y_{r+1})|\Y_r]\}$ is the conditional joint
entropy of $(\Theta,Y_{r+1})$ given the first order statistic,
averaged with respect to $f(\mathbf y_r)$.
The expected uncertainty change $\mathcal{B}(\Y_r; Y_{r+1})$
for prediction of $(r+1)$st order statistic is defined similarly.
These measures, which can be referred to as the information bridge
between the first and $(r+1)$st order statistics,
are invariant under linear transformations, but can be negative.
It can be shown that for any parent distribution $g(x|\theta)$
where $\theta$ is the scale parameter and any prior $f(\theta)$,
\begin{eqnarray*}
M(\Y_r; \Theta)&=& \mathcal{B}[\Y_r; (\Theta,Y_{r+1})] + \log \biggl(\frac
{n}{n-r} \biggr),
\\
M(\Y_r; Y_{r+1}) &=& \mathcal{B}(\Y_r; Y_{r+1}) + \log \biggl(\frac{n}{n-r} \biggr).
\end{eqnarray*}
Clearly, $\mathcal{B}(\cdot,\cdot|r,n) \to M(\cdot,\cdot|r,n)$ as
$\frac{r}{n} \to0$.
So, the quantity $\log (\frac{n}{n-r} )$
can be interpreted as the finite sample correction factor for the information.

\section{Conclusions} \label{CONCL}

This article is the first attempt to study the relationship between the
parameter and predictive
information measures, the analytical behavior of the predictive
information in terms of prior parameters
and the effects of conditional dependence between the observable
quantities on the Bayesian information measures. We provided analytical
results and
showed applications in some statistical and modeling problems.

The measure of information that sample provides about the parameter and
prediction jointly
led to some new insights about the marginal parameter and predictive
information measures.
For the case of conditionally independent observations, decompositions
of the joint information
revealed that the parameter
information is in fact the measure of information about the parameter
and prediction jointly.
This finding implies that all existing results about Lindley's
information are applicable
to the joint measure of parameter and predictive information. In particular,
the reference posterior and the optimal design that maximize the sample
information about the parameter
are also optimal solutions for the sample information about
the parameter and prediction jointly.
Yet another information decomposition revealed that predictive
information is a part of
the information that sample provides about the parameter.

We examined interplay between the information measures and the prior
and design parameters
for two general classes of models: the linear models for the normal
mean, and a broad subfamily
of the exponential family. A few applications showed the usefulness of
the information measures
and some insights were developed.
A proposition provided the optimal designs with respect to the parameter
(joint) information and predictive information measures for an ANOVA
type model.
The results include the minimum sample sizes required in terms of the
given prior variances
and the covariate vector for the prediction. Another
proposition provided the optimal prior variance allocation scheme with
respect to the parameter
(joint) information for collinear regression, which includes the minimum
prior variance required for the problem.
Examples for the linear and the exponential family models revealed that
the predictive information provided by the conditionally independent
sample is only
a small fraction of the parameter (joint) information and the gap
between the parameter and
predictive information measures grows rapidly with the sample size.
This finding indicates
that despite the importance of prediction in the Bayesian paradigm, the
parameter takes
the major share of the information provided by conditionally
independent samples.
An example examined the parameter information when the parameter of
interest is the vector of
means of two treatments and the predictive information of interest is
the weighted average (or contrast) between
outcomes of the two treatments. This example
revealed that the loss of information about the parameter under the
optimal design for
predictive information is much higher than the loss of predictive
information under the optimal design
for the parameter information. The parameter is the major shareholder
of the sample information
so its loss is more severe than the loss of predictive information
under suboptimal designs.

We have examined, for the first time, the role of conditional
dependence between observable quantities on the sample information
about the parameter and prediction. For a dependent sequence, the joint
parameter and predictive information
decomposes into the parameter information (Lindley's measure) and an
information measure
mapping the conditional dependence.
We provided more specific results for correlated variables whose
distributions can be transformed to normal
and for the order statistics without any distributional assumption.
For the normal sample, we compared the information measures for the
independent, the intraclass correlation
and serial correlation models. We showed that the parameter information
decreases and predictive information
increases with the correlation. However, the joint information decreases
in the correlation to a minimum point, which is determined by the prior
precision and sample size, and then increases.
For conditionally dependent sequences, the dominance of parameter
information that
was noted for the conditionally independent samples does not hold.
Since all information measures increase with the sample size, loss of
parameter information due to dependence
can be offset by taking larger samples.

Order statistics also provided a context for information analysis of
conditionally Markovian sequences.
Extension of a result on information properties of order statistics was
needed to show that
the Markovian dependence measure depends neither on the model for the
data, nor on the prior
distribution for the parameter. By this finding, the reference
posterior that maximizes the
sample information about the parameter retains its optimality according
to the joint parameter
and predictive information measure of the order statistics.
An example illustrated implication in terms of the optimal number of
failures to be
observed under Type II censoring.

\renewcommand{\thetable}{\arabic{table}}
\setcounter{table}{1}
\begin{table*}
\tablewidth=390pt
\caption{Classification of articles on Lindley's measure of sample
information about the parameter and its predictive version}\label{tab2}
\begin{tabular*}{390pt}{@{\extracolsep{4in minus 4in}}l@{}}
\hline
Parameter information:\\
\quad Likelihood model and design:\\
\quad
\begin{tabular}{l}
Lindley (\citeyear{1956Lindley}, \citeyear{1957Lindley}, \citeyear{1961Lindley}), Stone (\citeyear{1959Stone}), El-Sayyed (\citeyear{1969El}), Brooks
(\citeyear{1980Brooks}, \citeyear{1982Brooks}), \\
Smith and Verdinelli (\citeyear{1980Sm}), Turrero (\citeyear{1989Tu}), Barlow and Hsiung (\citeyear{1983Barlow}),
\\
Soofi (\citeyear{1988Soofi}, \citeyear{1990Soofi}), Ebrahimi and Soofi (\citeyear{1990Ebrahimi}), Carlin and Polson
(\citeyear{1991Carlin}), Verdinelli and Kadane (\citeyear{1992VerdinelliK}), \\
Polson (\citeyear{1992Polson}), Verdinelli (\citeyear{1992Verdinelli}), Parmigiani and Berry (\citeyear{1994Parmigiani}), Chaloner
and Verdinelli (\citeyear{1995Chaloner}), \\
Carota et al. (\citeyear{1996Carota}), Singpurwalla (\citeyear{1996Singpurwalla}), Yuan and Clarke (\citeyear{1999Yuan})\\
\end{tabular}
\\
Prior and posterior distributions:\\
\quad
\begin{tabular}{l}
Bernardo (\citeyear{1979aBernardo}, \citeyear{1979bBernardo}), Soofi (\citeyear{1988Soofi}, \citeyear{1990Soofi}), Ebrahimi and Soofi (\citeyear{1990Ebrahimi}),
Bernardo and Rueda (\citeyear{2002Bernardo}), \\
Bernardo (\citeyear{2005Bernardo})\\
\end{tabular}
\\
Predictive information \\
\quad Likelihood model and design:\\
\quad
\begin{tabular}{l}
San Martini and Spezzaferri (\citeyear{1984San}), Amaral and Dunsmore (\citeyear{1985Amaral}),
Verdinelli (\citeyear{1993Verdinelli}), \\
Verdinelli et al. (\citeyear{1993Verdinelli}), Chaloner and Verdinelli (\citeyear{1995Chaloner}), Singpurwalla
(\citeyear{1996Singpurwalla}) \\
\end{tabular}
\\
\hline
\end{tabular*}
\end{table*}

\begin{appendix}
\section*{Appendix}\label{app}

\subsection{Classification of Literature}

Table \ref{tab2} gives a classification of literature on the Bayesian
applications of mutual information.
Several authors have used information in various Bayesian contexts,
which are not listed in Table \ref{tab2};
examples include Aitchison (\citeyear{1975Aitchison}), Zellner (\citeyear{1977Zellner}, \citeyear{1988Zellner}), Geisser
(\citeyear{1993Geisser}), Keyes and Levy (\citeyear{1996Keyes}),
Ibrahim and Chen (\citeyear{2000Ibrahim}), Brown, George and Xu (\citeyear{2008Brown}).
Nicolae, {Meng} and {Kong} (\citeyear{2008Nicolae}) defined some measures of fraction of missing
information and have pointed
out connection between their measures and the entropy, stating that
``essentially all measures we presented
have entropy flavor.'' Measures of information
for nonparametric Bayesian data analysis are also available (M\"uller
and Quintana, \citeyear{2004Muller}).
Since our focus is on the mutual information, for example, Lindley's
measure and its predictive version,
we did not discuss other information measures.

\subsection[Proof of Proposition 1]{Proof of Proposition \protect\ref{ALLOCATN}}

(a) Noting that $\lambda_j=n_j, j=1,\ldots, p$, and
letting $n_1=n-\sum_{j=2}^p n_j$ in (\ref{MREG}) gives the
first-order conditions
\begin{eqnarray*}
\frac{\partial M(\Y; \Thet| Z, \eta, V_0)}{\partial n_j}
&=&\frac{1}{2} \biggl[\frac{v_{0j}}{\eta+ v_{0j} n_j} - \frac
{v_{01}}{\eta+ v_{01}n_1} \biggr]
\\
&=&0,\quad  j=2,\ldots,p.
\end{eqnarray*}
Solutions to this system give $n^*_j, j=2,\ldots,p$, in (\ref
{OPTNIPARA}) and $n^*_1$ is found
from $n^*_1=n-\sum_{j=2}^p n^*_j$. It can be verified by the
second-order conditions that the
solutions give the maximum.

(b)
Using $V_1= (\eta V_0^{-1}+Z^\prime Z )^{-1}$ in (\ref{MREGPRED}) gives
\begin{eqnarray*}
&&M(\Y;Y_\nu|\z_\nu, Z, \eta, V_0)
\\
 &&\quad=
\frac{1}{2} \log (\eta^{-1}\z_\nu^\prime V_0 \z_\nu+ 1 )
\\
&&\qquad{}- \frac{1}{2} \log \bigl( \z_\nu^\prime (\eta V_0^{-1}+Z^\prime Z
)^{-1} \z_\nu+1 \bigr).
\end{eqnarray*}
The first term does not depend on the design, so it is sufficient to minimize
\begin{eqnarray*}
h(n_1, \ldots, n_p)&=& \z_\nu^\prime (\eta V_0^{-1}+Z^\prime Z )^{-1}
\z_\nu
\\
&=& \sum_{j=1}^p \frac{v_{0j}z_j^2}{\eta+ v_{0j} n_j}
\end{eqnarray*}
subject to the constraint $\sum_{j=1}^p n_j=n$. Letting $n_1=n-\sum
_{j=2}^p n_j$ gives the first-order conditions
\begin{eqnarray*}
\frac{\partial h(n_1, \ldots, n_p)}{\partial n_j}
&=&-\frac{v_{0j}^2z_j^2}{(\eta+ v_{0j} n_j)^2} + \frac
{v_{01}^2z_1^2}{(\eta+ v_{01}n_1)^2}
\\
&=&0, \quad  j=2,\ldots,p.
\end{eqnarray*}
Solutions to this system give $n^*_j, j=2,\ldots,p$, in (\ref
{OPTNIPRED}) and $n^*_1$ is found
from $n^*_1=n-\sum_{j=2}^p n^*_j$. It can be verified by the
second-order conditions that the
solutions give the maximum.

\subsection[Proof of Proposition 2]{Proof of Proposition \protect\ref{ALLOCATV}}

The solutions are found similarly to part (a) of Proposition \ref{ALLOCATN} by
taking the derivative of (\ref{MREG})
with respect to $v_{0j}$ subject to $\sum_{j=1}^p v_{0j} =c$.

\subsection{Computation of Normal Predictive Correlation}

We compute the predictive correlation $\rho_p$ through the well-known
formula for partial correlation:
\renewcommand{\theequation}{A.\arabic{equation}}
\setcounter{equation}{0}
\begin{equation}
\label{RIJ|K}
\rho_{ij|k} = \frac{\rho_{ij}-\rho_{ik}\rho_{jk}}{(1-\rho
^2_{ik})^{1/2}(1-\rho^2_{jk})^{1/2}}.
\end{equation}
In our case, $i,j,k$ represent $Y_i, Y_\nu$ and $\theta$, respectively.
Note that
\[
\rho^2_{i\theta}=1-\frac{\sigma^2_{\theta|y_i}}{\sigma^2_0}=
\frac{1}{1+\eta}  \quad\mbox{for all }  i=1,2,\ldots .
\]
Letting $\rho^2_{ik}=\rho^2_{jk}=\rho^2_{i\theta}$ in (\ref
{RIJ|K}) gives the unconditional (predictive)
correlation as
\[
\rho_{i\nu}= \rho^2_{i\theta}+(1-\rho^2_{i\theta})\rho_{i\nu
|\theta}= \frac{1+\eta\rho_{i\nu|\theta}}{1+\eta}.
\]
Letting $\rho_{i\nu|\theta}=0, \rho, \rho_{\nu-i},  \nu> i$,
respectively for UC, IC and SC models,
we obtain the entries of Table \ref{tab1} for the three models.

\end{appendix}

\section*{Acknowledgments}

We thank the reviewers for their comments and suggestions which led us
to improve the exposition.
This research was instigated following response to a question about
usefulness of the prior predictive
distribution raised by Jie Feng in a doctoral seminar
course on Bayesian Statistics at Sheldon B. Lubar School of Business.
Ehsan Soofi's research was partially supported by
a Sheldon B. Lubar School's Business Advisory Council Summer Research
Fellowship.


\begin{thebibliography}{99}


\bibitem[\protect\citeauthoryear{}{1970}]{1970Abramowitz}
\textsc{Abramowitz}, M. and \textsc{Stegun}, I. A. (1970). \textit
{Handbook of Mathematical Functions, with Formulas,
and Mathematical Tables}. Dover, New York.

\bibitem[\protect\citeauthoryear{}{1975}]{1975Aitchison}
\textsc{Aitchison}, J. (1975). Goodness of prediction fit. \textit
{Biometrika} \textbf{62} 547--554.
\MR{0391353}

\bibitem[\protect\citeauthoryear{}{1985}]{1985Amaral}
\textsc{Amaral-Turkman}, M. A. and \textsc{Dunsmore}, I. (1985).
Measures of information in the predictive
distribution. In  \textit{Bayesian Statistics} (J. M.
Bernardo, M. H. DeGroot, D. V. Lindley
and A. F. M. Smith, eds.) \textbf{2} 603--612. North-Holland, Amsterdam.
\MR{0862505}

\bibitem[\protect\citeauthoryear{}{1992}]{1992Arnold}
\textsc{Arnold}, B., \textsc{Balakrishnan}, N. and \textsc{Nagaraja}, H. N. (1992).
 \textit{First Course in Order Statistics}. Wiley,
New York.
\MR{1178934}

\bibitem[\protect\citeauthoryear{}{1983}]{1983Barlow}
\textsc{Barlow}, R. E. and \textsc{Hsiung}, J. H. (1983). Expected
information from a life test experiment.
\textit{The Statistician} \textbf{48} 18--21.

\bibitem[\protect\citeauthoryear{}{1991}]{1991Belsley}
\textsc{Belsley}, D. A. (1991). \textit{Conditioning Diagnostics:
Collinearity and Weak Data in Regression}. Wiley, New York.
\MR{1090322}

\bibitem[\protect\citeauthoryear{}{2005}]{2005Bernardo}
\textsc{Bernardo}, J. M. (2005). Reference analysis. In \textit
{Handbook of Statistics}
(D. K. Dey and C. R. Rao, eds.) \textbf{25} 17--90. Elsevier, Amsterdam.
\MR{2490522}

\bibitem[\protect\citeauthoryear{}{2002}]{2002Bernardo}
\textsc{Bernardo}, J. M. and \textsc{Rueda}, R. (2002). Bayesian
hypothesis testing: A
reference approach. \textit{Int. Statist. Rev.} \textbf
{70} 351--372.

\bibitem[\protect\citeauthoryear{}{1979a}]{1979aBernardo}
\textsc{Bernardo}, J. M. (1979a). Expected information as expected
utility. \textit{Ann. Statist.} \textbf{7} 686--690.
\MR{0527503}

\bibitem[\protect\citeauthoryear{}{1979b}]{1979bBernardo}
\textsc{Bernardo}, J. M. (1979b). Reference posterior distribution for
Bayesian inference (with discussion).
\textit{J. Roy. Statist. Soc. Ser. B} \textbf{41} 605--647.
\MR{0547240}

\bibitem[\protect\citeauthoryear{}{1980}]{1980Brooks}
\textsc{Brooks}, R. J. (1980). On the relative efficiency of two
paired-data experiment.
\textit{J. Roy. Statist. Soc. Ser. B} \textbf{42} 186--191.
\MR{0583354}

\bibitem[\protect\citeauthoryear{}{1982}]{1982Brooks}
\textsc{Brooks}, R. J. (1982). On loss of information through
censoring. \textit{Biometrika} \textbf{69} 137--144.
\MR{0655678}

\bibitem[\protect\citeauthoryear{}{2008}]{2008Brown}
\textsc{Brown}, L. D., \textsc{George,} E. I. and \textsc{Xu}, X.
(2008). Admissible predictive density estimation.
\textit{Ann. Statist.} \textbf{36} 1156--1170.
\MR{2418653}

\bibitem[\protect\citeauthoryear{}{1995}]{1995Chaloner}
\textsc{Chaloner}, K. and \textsc{Verdinelli}, I. (1995). Bayesian
experimental design: A review.
\textit{Statist. Sci.} \textbf{10} 273--304.
\MR{1390519}

\bibitem[\protect\citeauthoryear{}{1991}]{1991Cover}
\textsc{Cover}, T. M. and \textsc{Thomas}, J. A. (1991). \textit
{Elements of Information Theory}. Wiley, New York.
\MR{1122806}

\bibitem[\protect\citeauthoryear{}{1991}]{1991Carlin}
\textsc{Carlin}, B. P. and \textsc{Polson}, N. G. (1991). An expected
utility approach to influence diagnostics.
\textit{J. Amer. Statist. Assoc.} \textbf{87} 1013--1021.

\bibitem[\protect\citeauthoryear{}{1996}]{1996Carota}
\textsc{Carota}, C., \textsc{Parmigiani}, G. and \textsc{Polson}, N.
G. (1996). Diagnostic
measures for model criticism. \textit{J. Amer. Statist. Assoc.}
\textbf{91} 753--762.
\MR{1395742}

\bibitem[\protect\citeauthoryear{}{1992}]{1992Ebrahimi}
\textsc{Ebrahimi}, N. (1992). Prediction intervals for future failures
in the exponential distribution under
hybrid censoring. \textit{IEEE Trans. Reliability} \textbf{41} 127--132.

\bibitem[\protect\citeauthoryear{}{1990}]{1990Ebrahimi}
\textsc{Ebrahimi}, N. and \textsc{Soofi}, E. S. (1990). Relative
information loss under type II censored
exponential data. \textit{Biometrika} \textbf{77} 429--435.
\MR{1064821}

\bibitem[\protect\citeauthoryear{}{2010}]{2010Ebrahimi}
\textsc{Ebrahimi}, N., \textsc{Soofi}, E. S. and \textsc{Soyer}, R.
(2010). Information measures in perspective.
\textit{Int. Statist. Rev.} \textbf{78}
doi:
\href{http://dx.doi.org/10.1111/j.1751-5823.2010.00105.x}{10.1111/j.1751-5823.2010.00105.x}.
To appear.

\bibitem[\protect\citeauthoryear{}{2004}]{2004Ebrahimi}
\textsc{Ebrahimi}, N., \textsc{Soofi}, E. S. and \textsc{Zahedi}, H.
(2004). Information properties of order statistics
and spacings. \textit{IEEE Trans. Inform. Theory} \textbf{50} 177--183.
\MR{2051426}

\bibitem[\protect\citeauthoryear{}{1969}]{1969El}
\textsc{El-Sayyed}, G. M. (1969). Information and sampling from
exponential distribution.
\textit{Technometrics} \textbf{11} 41--45.

\bibitem[\protect\citeauthoryear{}{1993}]{1993Geisser}
\textsc{Geisser}, S. (1993). \textit{Predictive Inference: An
Introduction}. Chapman and Hall, New York.
\MR{1252174}

\bibitem[\protect\citeauthoryear{}{1971}]{1971Good}
\textsc{Good}, I. J. (1971). Discussion of article by R. J. Buehler. In
\textit{Foundations of Statistical Inference} (V. P. Godambe and D. A.
Sprott, eds.) 337--339.
Holt, Rinehart and Winston, Toronto, ON.

\bibitem[\protect\citeauthoryear{}{2000}]{2000Ibrahim}
\textsc{Ibrahim}, J. G. and \textsc{Chen}, M. H. (2000). Power prior
distributions for regression models.
\textit{Statist. Sci.} \textbf{15} 46--60.
\MR{1842236}

\bibitem[\protect\citeauthoryear{}{1985}]{1985Kaminsky}
\textsc{Kaminsky}, K. S. and \textsc{Rhodin}, L. S. (1985). Maximum
likelihood prediction.
\textit{Ann. Inst. Statist. Math.} \textbf{37} 505--517.
\MR{0818048}

\bibitem[\protect\citeauthoryear{}{1996}]{1996Keyes}
\textsc{Keyes}, T. K. and \textsc{Levy}, M. S. (1996). Goodness of
prediction fit for multivariate linear models.
\textit{J. Amer. Statist. Assoc.} \textbf{91} 191--197.
\MR{1394073}

\bibitem[\protect\citeauthoryear{}{1959}]{1959Kullback}
\textsc{Kullback}, S. (1959). \textit{Information Theory and
Statistics}. Wiley, New York
(reprinted in 1968 by Dover).
\MR{0103557}

\bibitem[\protect\citeauthoryear{}{1971}]{1971Lawless}
\textsc{Lawless}, J. L. (1971). A prediction problem concerning
samples from the exponential distribution
with application in life testing. \textit{Technometrics} \textbf{13} 725--730.

\bibitem[\protect\citeauthoryear{}{1956}]{1956Lindley}
\textsc{Lindley}, D. V. (1956). On a measure of information provided
by an
experiment. \textit{Ann. Math. Statist.} \textbf{27} 986--1005.
\MR{0083936}

\bibitem[\protect\citeauthoryear{}{1957}]{1957Lindley}
\textsc{Lindley}, D. V. (1957). Binomial sampling schemes and the
concept of information.
\textit{Biometrika} \textbf{44} 179--186.
\MR{0087273}

\bibitem[\protect\citeauthoryear{}{1961}]{1961Lindley}
\textsc{Lindley}, D. V. (1961). The use of prior probability
distributions in statistical
inference and decision. In \textit{Proceedings of the Fourth Berkeley
Symposium Math. Statist. Probab.}
(J. Neyman, ed.) \textbf{1} 436--468. Univ. California Press, Berkeley,
CA.
\MR{0156437}

\bibitem[\protect\citeauthoryear{}{2008}]{2008Nicolae}
\textsc{Nicolae}, D. L., \textsc{Meng}, X.-L. and \textsc{Kong}, A.
(2008). Quantifying the fraction of missing
information for hypothesis testing in statistical and genetic studies
(with discussion).
\textit{Statist. Sci.} \textbf{23} 287--331.
\MR{2483902}

\bibitem[\protect\citeauthoryear{}{2010}]{2010Maruyama}
\textsc{Maruyama}, Y. and \textsc{George}, E. I. (2010). Fully Bayes
model selection with a generalized $g$-prior.
Working paper. Univ. Pennsylvania.

\bibitem[\protect\citeauthoryear{}{2004}]{2004Muller}
\textsc{M\"uller}, P. and \textsc{Quintana}, F. A. (2004).
Nonparametric Bayesian data analysis.
\textit{Statist. Sci.} \textbf{23} 287--331.
\MR{2082149}

\bibitem[\protect\citeauthoryear{}{1994}]{1994Parmigiani}
\textsc{Parmigiani}, G. and \textsc{Berry}, D. A. (1994). Applications
of Lindley information measures
to the design of clinical experiments. In \textit{Aspects of
Uncertainty: A Tribute to D. V. Lindley}
(P. R. Freeman and A. F. M. Smith, eds.) 329--348. Wiley, Chichester, UK.
\MR{1309700}

\bibitem[\protect\citeauthoryear{}{1992}]{1992Polson}
\textsc{Polson}, N. G. (1992). On the expected amount of information
from a nonlinear model.
\textit{J. Roy. Statist. Soc. Ser. B} \textbf{54} 889--895.
\MR{1185230}\

\bibitem[\protect\citeauthoryear{}{1993}]{1993Polson}
\textsc{Polson}, N. G. (1993). A Bayesian perspective on the design of
accelerated life tests.
In \textit{Advances in Reliability} (A. P. Basu, ed.) 321--330. North-Holland, Amsterdam.

\bibitem[\protect\citeauthoryear{}{2000}]{2000Pourahmadi}
\textsc{Pourahmadi}, M. and \textsc{Soofi}, E. S. (2000). Predictive
variance and information worth of observations
in time series. \textit{J. Time Ser. Anal.} \textbf{21} 413--434.
\MR{1787663}

\bibitem[\protect\citeauthoryear{}{1984}]{1984San}
\textsc{San Martini}, A. and \textsc{Spezzaferri}, F. (1984). A
predictive model selection criteria.
\textit{J. Roy. Statist. Soc. Ser. B} \textbf{46} 296--303.
\MR{0781890}

\bibitem[\protect\citeauthoryear{}{1948}]{1948Shannon}
\textsc{Shannon}, C. E. (1948). A mathematical theory of communication.
\textit{Bell Syst. Tech. J.} \textbf{27} 379--423.
\MR{0026286}

\bibitem[\protect\citeauthoryear{}{1996}]{1996Singpurwalla}
\textsc{Singpurwalla}, N. D. (1996). Entropy and information in reliability.
In \textit{Bayesian Analysis of Statistics and Econometrics: Essays in
Honor of Arnold Zellner}
(D. Berry, K. Chaloner and J.~Geweke, eds.) 459--469. Wiley, New
York.
\MR{1367335}

\bibitem[\protect\citeauthoryear{}{1980}]{1980Sm}
\textsc{Smith, A. F. M.} and \textsc{Verdinelli}, I. (1980). A note on Bayesian design for
inference using ahierarchical linear model. \textit{Biometrika} \textbf{67}
613--619.

\bibitem[\protect\citeauthoryear{}{1988}]{1988Soofi}
\textsc{Soofi}, E. S. (1988). Principal component regression under
exchangeability.
\textit{Comm. Statist. Theory and Methods} \textbf{A17} 1717--1733.
\MR{0945783}

\bibitem[\protect\citeauthoryear{}{1990}]{1990Soofi}
\textsc{Soofi}, E. S. (1990). Effects of collinearity on information
about regression coefficients.
\textit{J. Econometrics} \textbf{43} 255--274.\break
\MR{1046959}

\bibitem[\protect\citeauthoryear{}{1987}]{1987Stewart}
\textsc{Stewart}, G. W. (1987). On collinearity and least squares
regression (with discussion).
\textit{Statist. Sci.} \textbf{2} 68--100.
\MR{0896260}

\bibitem[\protect\citeauthoryear{}{1959}]{1959Stone}
\textsc{Stone}, M. (1959). Application of a measure of information to the
design and comparison of regression experiments. \textit{Ann. Math.
Statist.} \textbf{29} 55--70.
\MR{0106528}

\bibitem[\protect\citeauthoryear{}{1989}]{1989Tu}
\textsc{Turrero}, A. (1989). On the relative efficiency of grouped and censored
survival data. \textit{Biometrika} \textbf{76} 125--131.\

\bibitem[\protect\citeauthoryear{}{1992}]{1992Verdinelli}
\textsc{Verdinelli}, I. (1992). Advances in Bayesian experimental
design. In \textit{Bayesian Statistics}
(J. O. Berger, J. M. Bernardo, A. P. Dawid and A. F. M. Smith, eds.)
 \textbf{4} 467--481. Wiley, New York.
\MR{1380292}

\bibitem[\protect\citeauthoryear{}{1992}]{1992VerdinelliK}
\textsc{Verdinelli}, I. and \textsc{Kadane}, J. B. (1992). Bayesian
designs for maximizing information and outcome.
\textit{J. Amer. Statist. Assoc.} \textbf{87} 510--515.
\MR{1173814}

\bibitem[\protect\citeauthoryear{}{1993}]{1993Verdinelli}
\textsc{Verdinelli}, I., \textsc{Polson}, N. G. and \textsc{Singpurwalla}, N. D. (1993).
Shannon information and Bayesian design for prediction in accelerated
life-testing.
In \textit{Reliability and Decision Making} (R. E. Barlow, C. A.
Clarotti and F. Spizzichino, eds.) 247--256.
Chapman and Hall, London.
\MR{1296278}

\bibitem[\protect\citeauthoryear{}{2003}]{2003West}
\textsc{West}, M. (2003). Bayesian factor regression models in the
``large $p$, small $n$'' paradigm.
In \textit{Bayesian Statist.} (J. M. Bernardo, M.~J.
Bayarri, J. O.
Berger, A. P. Dawid, D. Heckerman, A. F. M. Smith and M. West, eds.)
 \textbf{7} 723--732. Oxford Univ. Press.
\MR{2003537}

\bibitem[\protect\citeauthoryear{}{1999}]{1999Yuan}
\textsc{Yuan}, A. and \textsc{Clarke}, B. (1999). An information
criterion for likelihood selection.
\textit{IEEE Trans. Inform. Theory} \textbf{45} 562--571.
\MR{1677018}

\bibitem[\protect\citeauthoryear{}{1986}]{1986Zellner}
\textsc{Zellner}, A. (1986). On assessing prior distributions and
Bayesian regression analysis with $g$-prior
distributions. In \textit{Bayesian Inference and Decision Techniques}
(P. Goel and A. Zellner, eds.)
233--243. North-Holland, Amsterdam.
\MR{0881437}

\bibitem[\protect\citeauthoryear{}{1977}]{1977Zellner}
\textsc{Zellner}, A. (1977). Maximal data information prior distributions.
In \textit{New Developments in the Applications of Bayesian Methods}
(A. Aykac and C. Brumat, eds.) 211--232. North-Holland, Amsterdam.
\MR{0505477}

\bibitem[\protect\citeauthoryear{}{1988}]{1988Zellner}
\textsc{Zellner}, A. (1988). Optimal information processing and Bayes'
theorem (with discussion). \textit{Amer. Statist.} \textbf{42} 278--284.\vspace*{-1pt}
\MR{0971095}

\end{thebibliography}
\end{document}